\documentclass[prd,aps,nofootinbib,nobibnotes,superscriptaddress,preprintnumbers]{revtex4}

\usepackage{color}
\usepackage{amsmath}
\usepackage{amsmath}
\usepackage{amsfonts}
\usepackage{graphicx}
\usepackage{slashed}
\usepackage[dvipsnames]{xcolor}

\usepackage{hyperref}
\usepackage{float}

\usepackage{physics}

\newcommand{\be}{\begin{equation}}
\newcommand{\ee}{\end{equation}}

\newcommand{\nb}{{\omega_c^{-1}}}
\newcommand{\cL}{{\cal{L}}}
\newcommand{\cF}{{\cal{F}}}
\newcommand{\tw}{\tilde{\omega}}
\newcommand{\pgw}{{\cal{P}}_{\rm GW}}
\newcommand{\Geff}{G_N^{\rm eff}}
\newcommand{\cQ}{{\cal{Q}}}

\definecolor{tabblue}{HTML}{1f77b4}
\definecolor{taborange}{HTML}{ff7f0e}
\definecolor{tabgreen}{HTML}{2ca02c}
\definecolor{tabred}{HTML}{d62728}
\definecolor{tabpurple}{HTML}{9467bd}

\begin{document}

\title{Scalar kicks and memory}

\author{Philippe Brax}
\affiliation{Institut de Physique Th\'eorique, Universit\'e Paris-Saclay,CEA, CNRS, F-91191 Gif-sur-Yvette Cedex, France.}

\author{Dani\`ele A.~Steer}
\affiliation{Laboratoire de Physique de l’\'Ecole Normale Sup\'erieure, ENS, CNRS, Universit\'e PSL, Sorbonne Universit\'e, Universit\'e Paris Cit\'e, F-75005 Paris, France}
\date{\today}

\begin{abstract}
     A scalar field coupled conformally and disformally to matter affects both the linear memory effect for binary systems on hyperbolic orbits, as well as the kick velocity for binaries on bound or unbound orbits.  We study these corrections in detail, their order of magnitude, and discuss their detectability. In particular, we find that the disformal interaction does not contribute to the memory effect and the emitted power spectrum at zero frequency. The conformal interaction corrects the GR linear memory and the quadrupole emitted power at zero frequency resulting in a breaking of the GR memory-power spectrum relationship. On the other hand, disformal interactions give rise to a change of momentum for the centre of mass. Hence, measuring both the linear memory effect and the kicks for hyperbolic orbits would give access to the conformal and disformal couplings of nearly massless scalars to matter.
\end{abstract}

\maketitle

\section{Introduction}
Light scalar fields could play many roles in cosmology, for instance they could be part of the dark matter of the Universe \cite{Hui:2021tkt} or trigger dynamical dark energy \cite{Copeland:2006wr}. Unless protected by a yet unknown symmetry, light scalar fields \cite{Brax:2021wcv} should also couple to matter\footnote{Their coupling to matter is always at least mediated by gravity to which both the scalars and matter are coupled. }. These couplings are problematic in the solar system where tests of gravitational interactions are very precise, ranging from the Cassini probe of the Shapiro delay \cite{Bertotti:2003rm} to the Microscope test of the equivalence principle \cite{Berge:2017ovy,MICROSCOPE:2022doy}. They could also be a blessing as they could open a wealth of new phenomena to observations, for instance in the case of binary systems where scalars could be radiated and back react on the orbits of the system \cite{BraxKuntz}. 

Here
we mainly (but not exclusively) consider hyperbolic orbits (scattering) in scalar-tensor theories of gravity with conformal and disformal couplings. {These theories are metric in the sense that matter couples to a single metric \cite{Will:2014kxa}, the Jordan metric, differing from the Einstein metric defining the Einstein-Hilbert term. The equivalence principle is therefore respected as all species are universally coupled to gravity. Our choice of Jordan metric depending on a single massless scalar field and involving both a conformal and a disformal coupling makes the models that we consider part of the Hornesdki family \cite{Bettoni:2013diz,Sakstein:2015jca,Kobayashi:2019hrl}. } In the conservative sector and working to 1PN, we show how the precession of the orbits is affected by these couplings, see also \cite{Brax:2018bow,Benisty:2022lox,Brax:2020vgg}. In the radiative sector, we focus on two phenomena:
the kick velocity and the linear memory effect. In General Relativity (GR), the anisotropic radiation of gravitational waves (GWs) means that linear momentum is lost from the system resulting in the recoil or ``kick'' of the centre of mass position.\footnote{See \cite{LIGOScientific:2020ufj} for LVK constraints on the remnant kick velocity GW190521, and \cite{Ranjan:2024wui} who investigate the possibility of precise measurements of the kick velocity using multiband GW networks (LISA, LVK and 3G detectors such as ET/CE).} For hyperbolic encounters of two compact objects with different masses and high eccentricity $e>1$ and spins, the kick velocities can be as large as 10000 km/s \cite{PhysRevLett.102.041101}. In modified gravity  scalar radiation emitted near to the point of closest approach will modify the kick, as we discuss for non-spinning bodies. For compact objects such as neutron stars (NS), known bounds on the disformal suppression scale \cite{Brax:2014vva,Brax:2015hma}  prevent any significant effect. For pairs of White Dwarfs (WD), however, no bound on the disformal coupling is known, and the kick velocity might be detectable. This can only happen if the disformal coupling scale $\Lambda$ depends on the environment and differs for WD from their NS counterparts.  In this paper, we are agnostic and consider that the scalar field theories are effective field theories whose coupling constants are specific to a particular experimental setup, i.e.~probing different energy scales such that the nuclear densities of NS or the less dense WD might result in different suppression scales $\Lambda$.
The dimensionless ratio between the disformal effects and GR is governed by 
\be 
\epsilon_\Lambda=  \frac{\beta^2 G_N m}{ \Lambda^2 p^3}
\ee
where $\beta$ is the conformal coupling of the scalar to matter, $m$ is the total mass of the binary system, and $p$ the typical size of the orbit.

We are also interested in gravitational wave memory effects occurring when there is a permanent change $\Delta h_{ij}^{\rm{TT}}$ in the GW form. {Here the metric property of the theories we consider is crucial. Indeed this guarantees that memory depends on the variation of the unique Jordan metric coupling gravity to matter.} The simplest memory effect, the linear one, takes place when there is net change in the third time
derivatives of multipole moments of the system, e.g.~during hyperbolic trajectories or
an asymmetric supernova explosion see e.g.~\cite{Mukhopadhyay:2021zbt}.  In general, there is a resulting permanent displacement of test bodies whose potential detection with is under investigation, see \cite{NANOGrav:2023vfo} for PTAs and e.g.~\cite{Inchauspe:2024ibs,Goncharov:2023woe,Gasparotto:2023fcg} for LISA. We investigate the linear memory effect when coupled scalars are present and determine the  contributions from the disformally coupled fields. 
Here we find that while the scalar kicks are dictated by $\epsilon_\Lambda$, the memory effect is {\it independent} of the disformal coupling depending only on {the conformal coupling $\beta$.}
Thus, if ever experiments became precise enough to test kicks, memory effects and the power spectrum of radiations from hyperbolic binary systems, their observation would provide a convincing manner of disentangling the couplings of scalars to matter.  {To put this work into context, note that gravitational memory effects have been studied in \cite{Heisenberg:2023prj} for Horndeski models. The  influence of the conformal coupling taken in the form of Brans-Dicke theory, as well as the role of asymptotic symmetries is discussed in \cite{Tahura:2020vsa,Hou:2020xme}. The particular role played by a dual two-form field to the scalar field,  in the link  between memory and asymptotic symmetries in Brans-Dicke theories can be found in \cite{Seraj:2021qja}. }

This paper is setup as follows. In section \ref{sec:GReffectsGalore}, working in GR, we review effects related to hyperbolic orbits: precession, the linear memory effect, and kicks, which are all well known. 
The linear memory effect is a very old problem \cite{1987Natur.327..123B}, but despite that even the recent literature is not always clear and updates/corrections to different calculations exist, particularly regarding the GW energy spectrum $\pgw(\omega)$ at zero frequency $\omega=0$ (directly related to the linear memory effect). Then, in the remaining sections of the paper we add a massless scalar with disformal coupling to matter.   The model is summarized in section \ref{sec:scalar-precession} where we focus on the conservative dynamics and work with the 1PN Einstein-Infeld-Hoffman action to 1PN including disformal effects, deriving the equations of motion in the centre of mass frame. This then leads on to the calculation of the precession of hyperbolic orbits with a scalar field. In the remainder of the paper we consider the radiative sector. In section \ref{sec:scalar-memory} we determine the contribution of the scalar field to the  memory effect, and determine the scalar Jordan displacement from a hyperbolic system. Scalar kicks are the subject of section \ref{sec:scalar-kicks}, and our conclusions are given in section \ref{sec:conc}.

\section{Hyperbolic Effects  in GR}
\label{sec:GReffectsGalore}

We initially summarise well known results on hyperbolic orbits in GR: precession, the linear memory effect and the kicks due to the emission of gravitational waves.  In the remainder of the paper these effects will be revisited for scalar-tensor theories with disformal couplings.  

Our basic notation is the following. We consider non-spinning binary systems of point-particles $A,B$ in the $(x,y)$ plane, with masses $m_{A,B}$, relative position $\vec{r} = \vec{x}_A - \vec{x}_B$ and relative velocity $\vec{v} = \vec{v}_A - \vec{v}_B$. These define the two unit vectors $\vec{n}$ and $\vec{\lambda}$ through
\begin{eqnarray}
     \vec{r} = r \vec{n}, &\qquad & \vec{n} = (\cos\phi,\sin\phi,0)
    \label{eq:vecr}
    \\
    \vec{v} = \dot{r}\vec{n} + r \dot{\phi} \vec{\lambda} ,&\qquad& \vec{\lambda} = (-\sin\phi,\cos\phi,0).
    \label{eq:vecv}
\end{eqnarray}
The observer/detector will be at position $\vec{R}=R\vec{N}$ where $\vec{N}$ is a unit vector. The 0th order Newtonian equations are given in Appendix \ref{app:Newton}, and in the centre of mass (CM) frame they read
\begin{eqnarray}
    r(\phi) &=& \frac{p}{1+e\cos(\phi)},
    \label{eq:r}
    \\
    \dot{\phi} &=& \sqrt{\frac{G_N m}{p^3}} (1+e\cos(\phi))^2
    \label{eq:dotp}
\end{eqnarray}
where $e$ is the eccentricity, $m$ the total mass
\be
m=m_{A} + m_B,
\ee
and the periastron $p$ defines the closest approach $r_{\rm min} =p/(1+e)$.  The total conserved orbital energy is $E=\nu \frac{Gm^2}{2p}(e^2-1)$ with the mass ratio 
\be
\nu= \frac{m_A m_B}{m^2}.
\ee
Bound elliptical orbits have $e<1$ and $0\leq \phi < 2\pi$. On unbound hyperbolic orbits $e>1$ with $\phi_-(e)\leq  \phi < \phi_+(e)$ where $\phi_\pm = \pm \cos^{-1}(1/e)$, and as $\phi \rightarrow \phi_\pm$, $v \rightarrow v_\infty$ where 
\begin{equation}
    v_\infty^2 = \frac{G_N m}{p}(e^2-1).
    \label{eq:vinfty}
\end{equation}

\subsection{Precession}
\label{subsec:precession}

While angular momentum and energy conservation imply that there are no secular changes to either $e$ and $p$ to first post-Newtonian (PN) order in GR, both hyperbolic and elliptical orbits precess \cite{poisson_will_2014}.
Following
\cite{Damour1985}, namely on rewriting the Newtonian orbit in the ``quasi-Newtonian'' form 
\begin{equation}
    r = \frac{p}{1+e\cos f}
    \nonumber
\end{equation}
where $f=\phi-w$ (with $w=0$ in the Newtonian limit), then to 1st PN order $w$ is given by (we reinstate factors of $c$) \cite{poisson_will_2014}
\begin{equation}
    w=\frac{1}{e}\frac{G_N m}{c^2 p} \int df \left\{ 
    3e - \left[3-\nu - \frac{1}{8}(8+21\nu)e^2\right]\cos f - (5-4\nu)e\cos2f + \frac{3}{8}\nu e^2 \cos 3f
    \right\}.
\end{equation}
For elliptical orbits ($0<f<2\pi$), 
only the first term contributes giving the standard text-book expression for the precession 
\begin{equation}
    w_{\rm e}= 2 \pi \Delta^{\rm{GR}}
\end{equation}
with
\begin{eqnarray}
    \Delta^{\rm{GR}} \equiv  \frac{3G_N m}{c^2 p}.
    \label{eq:DeltaGR}
\end{eqnarray}
For hyperbolic orbits $\phi_- \leq f < \phi_+$ and
\begin{equation}
 w_{\rm h} = 
\Delta^{\rm{GR}} \left\{ 2 \arccos(-1/e)
     +
    \frac{\sqrt{e^2-1}}{3 e^2}  \left[
     2(2+e^2) + 5\nu(e^2-1) 
     \right] \right\}
 \end{equation}
 which differs from the expression for elliptical orbits\footnote{The results of \cite{Caldarola:2023ipo} assume $w_e$ for hyperbolic orbits.}  though of course they agree when $e=1$.
{For large eccentricity $w_{\rm h}$ increases linearly with $e$, $w_{\rm h} \sim e \Delta^{\rm{GR}}  (2+5\nu)/3$}. In section \ref{sec:prec} we determine the modifications to these expressions for disformally coupled scalars.

\subsection{GWs and the linear memory effect}
\label{subsec:GWmemory}

\subsubsection{Linear memory and energy spectrum} 
\label{subsec:linmem}

The linear memory effect is a non-zero change in the 3D transverse and traceless (TT) metric perturbation between $t=\pm \infty$, and leads to a permanent displacement $\delta L/L$ in the arm lengths of a GW interferometer. 
In the quadrupole approximation (see e.g.~\cite{Maggiore:2007ulw})
\be
h_{ij}(t,\vec{R}) = \frac{2G_N}{c^4 R} \Lambda_{ij,k\ell}(\vec{N}) \ddot{I}^{k\ell}(t_R)
\label{eq:basic}
\ee
where $R$ is the distance to the source at position $\vec{R}=R\vec{N}$, $\Lambda_{ij,k\ell}(\vec{N})$ is the projection tensor onto TT components, and ${I}^{k\ell}$ is the quadrupole moment evaluated at the retarded time $t_R=t-R/c$. 
The variation of the metric between $t=
\pm \infty$ is given by
\be
\Delta h_{ij}= \int_{-\infty}^{\infty} {\rm d}t \, \dot h_{ij}(t).
\ee
Substituting \eqref{eq:basic}
and working in Fourier space (with convention $\tilde{I}_{ij} (\omega)=\int dt I_{ij}(t) e^{-i\omega t}$) leads to
\be
\Delta h_{ij} = -i \frac{2G_N }{c^4 R} \Lambda_{ij,k\ell} \left. \left[\omega^3 \tilde{I}^{k\ell}(\omega) \right] \right|_{\omega=0}.
\label{eq:nlm}
\ee
Thus a non-zero linear memory effect, $\Delta h_{ij} \neq 0$ implies $\lim_{\omega \rightarrow 0} \left[ \omega^3 I^{k\ell}(\omega)\right] \neq 0$.  Correspondingly 
the permanent change in the invariant 
distance between two test masses is 
\be 
\delta^{(2)}= \frac{1}{2}\Delta h_{ij}\hat{\ell}^i \hat{\ell}^i
\label{eq:memoryGRlength}
\ee
where $\hat{\ell}^i$ is the unit vector between the test masses.

Alternatively $\Delta h_{ij}$ can be rewritten in terms of the energy spectrum $\pgw(\omega)$ of the emitted GWs. In the quadrupole approximation, the total energy emitted in GWs is
\be 
E_{\rm GW}= \frac{G_N }{5c^5} \int_{-\infty}^{\infty} dt (\dddot I_{ij})^2
=\frac{G_N  }{5 \pi c^5} \int_0^{\infty} {d} \omega \, \omega^6 \vert I_{ij}(\omega)\vert^2
\ee
so that
\be
\pgw(\omega) = \frac{G_N}{5 \pi c^5} \left[ \omega^3 \tilde{I}_{ij}(\omega)\right]\left[ \omega^3 \tilde{I}_{ij}^*(\omega) \right].
\label{eq:Pomega}
\ee
Thus comparing with Eq.~\eqref{eq:nlm}, a non-vanishing linear memory effect corresponds to $\pgw(0)\neq 0$.

\subsubsection{Linear memory effects for hyperbolic orbits }
\label{subsec:memory}

We now focus on binary system whose orbits are hyperbolic. Calculation of $\pgw(\omega)$ (see below) shows that a short GW burst of characteristic frequency $$
\omega_{\rm max}\sim 6\sqrt{\frac{G_N m}{r_{\rm min}^3}}\frac{\sqrt{e-1}}{e+1}
$$ is emitted near $r_{\rm min}$ with a corresponding linear memory effect $\Delta h_{ij} \neq 0$. Depending on $m, e$ and $r_{\rm min}\gg r_s\equiv 2Gm$, this characteristic frequency $\omega_{\rm max}$ may fall in the detection band of different experiments, and be detectable if their amplitude is large enough. Using the O3b data of LVK, the corresponding  density rate of such events is reported to be $\lesssim 10^{-4}$ per year per Gpc$^3$ \cite{Bini:2023gaj}. 
The result of many overlapping and unresolved bursts from hyperbolic orbits will also generate a stochastic GW background as whose characteristics and detectability (which depends on the expected number of hyperbolic orbit encounters) have been explored in \cite{Garcia-Bellido:2021jlq,Kerachian:2023gsa,Hait:2022ukn}.

We now calculate the energy spectrum for a single burst, $\pgw(\omega)$ given in Eq.~\eqref{eq:Pomega}, from first principles for convenience. This leading order calculation has been undertaken many times in the literature, see e.g.~\cite{Grobner:2020fnb,Caldarola:2023ipo,poisson_will_2014,1989MNRAS.239..845B} and references within. Despite that, some aspects of it are sometimes obscured by the formalism\footnote{There are also some incorrect statements: for instance as explained above $\pgw(0) \neq 0$ due to the linear memory effect. Yet some papers (also studying the linear memory effect) e.g.~\cite{DeVittori:2012da} have $\pgw(0)=0$}, and hence we highlight the salient features before extending the calculation to scalar-tensor theories in later sections.

We first calculate $\pgw(\omega=0)$.
For binary systems
the (traceless) quadrupole moment is 
given by
\be 
I^{ij} = \mu \left(r^ir^j-\frac{\delta^{ij}}{3} r^2\right)
\label{eq:Idefreal}
\ee
where $\mu=m_1 m_2/m$ is the reduced mass. It follows from Newton's equations that 
 as $t\rightarrow \pm\infty$, $\ddot{I}_{ij} \rightarrow 2\mu v_i v_j$. Furthermore, from Kepler's laws
(Appendix \ref{app:Newton})
\begin{equation}
    \vec{v} \underset{t\rightarrow \pm \infty}{\longrightarrow} \frac{v_\infty}{e} \left( 1, \pm {\sqrt{e^2-1}},0\right)
\end{equation}
where $v_\infty$ is given in Eq.~(\ref{eq:vinfty}). Hence from Eq.~\eqref{eq:basic}, 
\begin{eqnarray}
    \Delta h_{ij} &=& h_{ij}^+ - h_{ij}^-
    \nonumber
    \\
    &=& \frac{2G_N}{c^4 R} \Lambda_{ij,k\ell}(\vec{N}) (\ddot{I}_{k\ell}^+ - \ddot{I}_{k\ell}^-)
        \nonumber
    \\
    &=& -\frac{4G_N\mu }{c^4 R}  v_\infty^2 \frac{\sqrt{e^2-1}}{e^2} \Lambda_{ij,k\ell}(\vec{N})(\delta_{k,x}\delta_{\ell,y} + \delta_{\ell,x}\delta_{k,y}).
    \label{eq:realsp}
\end{eqnarray}
Thus from Eq.~\eqref{eq:memoryGRlength}
the permanent shift in the length of a rod pointing in the $\hat{\ell}^i$ direction is proportional to $G_N\mu v_\infty^2/R$ and reads
\be 
\delta^{(2)}_{\rm{GR}}=-\frac{4G_N\mu }{c^4 R}  v_\infty^2 \frac{\sqrt{e^2-1}}{e^2} \left[ \hat{\ell}_x \hat{\ell}_y -(\vec{N}\cdot\hat{\ell})(N_x\hat{\ell}_y+N_y \hat{\ell}_x)+ \frac{1}{2}\left((\vec{N}\cdot\hat{\ell})^2+1\right)N_xN_y\right].
\label{eq:delta2GR}
\ee
We will compare this expression to the scalar case below.  Furthermore, comparison with Eq.~\eqref{eq:nlm} gives 
\begin{equation}
     \underset{\omega \rightarrow 0}{\lim }   \left|\omega^3 \tilde{I}^{12}(\omega) \right| =
     \underset{\omega \rightarrow 0}{\lim }  \left|\omega^3 \tilde{I}^{21}(\omega)  \right| =  \frac{4G_N m\mu}{p}\frac{(e^2-1)^{3/2} }{e^2}
   \label{eq:I120}
\end{equation}
with all other components of $\tilde{I}^{ij}$ vanishing in the zero frequency limit.  Thus from Eq.~\eqref{eq:Pomega}, 
\begin{eqnarray}
    \pgw(0)= \frac{32 G_N \mu^2 }{5\pi c^5} \left(\frac{G_N m}{p} \right)^2 \frac{(e^2-1)^{3} }{e^4}=\frac{32 G_N \mu^2 }{5\pi c^5} \frac{v_\infty^2}{c^2} \frac{(e^2-1)^{} }{e^4}.
    \label{eq:P0}
\end{eqnarray}
This implies that there is an intrinsic relationship between 
the linear memory effect and the power spectrum at zero frequency
\be 
\frac{\delta^{(2)}_{\rm{GR}}}{\pgw(0)}=-\frac{5 c^3 }{ 8\mu R}   \frac{ e^2 }{\sqrt{e^2-1}} \left[ \hat{\ell}_x \hat{\ell}_y -(\vec{N}\cdot\hat{\ell})(N_x\hat{\ell}_y+N_y \hat{\ell}_x)+ \frac{1}{2}\left((\vec{N}\cdot\hat{\ell})^2+1\right)N_xN_y\right],
\label{eq:delta2GRbis}
\ee
which does not depend on the velocity of the binary system and only on intrinsic feature such as the eccentricity and the reduced mass together with the distance of the observer to the binary system's centre of mass. We will see below how this relation is disturbed by the presence of a scalar field.

To determine the energy spectrum $\pgw(\omega)$ for all $\omega$ using Eq.~(\ref{eq:Pomega}) we need to evaluate $\tilde{I}^{k\ell}(\omega)$, starting from Eq.~\eqref{eq:Idefreal}. Its calculation is simplified by introducing the mean anomaly $\eta$ defined by
\begin{equation}
\tanh (\eta/2)= \sqrt{\frac{e-1}{e+1}}   \tan(\phi/2)
\label{eq:etadef}
\end{equation}
for hyperbolic orbits, in terms of which Newtonian dynamics take the form (see Appendix \ref{app:Newton}) 
\begin{eqnarray}
     r(\eta) = a(e\cosh \eta - 1)
   &  \text{with}&   a = \frac{p}{e^2-1}
   \label{eq:reta}
   \\
      t(\eta) = \nb (e\sinh \eta - \eta) &
    \text{with}& \omega_c =\sqrt{\frac{G_Nm}{a^3}} , \label{eq:teta}
   \\
   r(\eta) \vec{n} = a\vec{m}  &  \text{with}&  \vec{m}= \big( e - \cosh\eta, \sqrt{e^2-1} \sinh\eta ,0\big).
    \label{eq:mdef}
\end{eqnarray}
Then Eq.~\eqref{eq:Idefreal} leads to
\begin{eqnarray}
    \tilde{I}^{k\ell}(\omega) &=& 
 \mu  \int dt \, r^2(t) \left(n^k(t) n^\ell(t) - \frac{\delta ^{k\ell}}{3}\right)e^{-i\omega t}
\nonumber
\\
 &=&   \left(\frac{\mu  a^2}{3\omega_c} \right) \int_{-\infty}^\infty {\rm d}\eta \; \left\{ \left[ 3m^k(\eta)m^\ell(\eta) - (e\cosh \eta - 1)^2 \delta^{k\ell} \right](e\cosh \eta - 1) \right\} e^{-i \tw(e\sinh\eta - \eta)}
    \label{here}
\end{eqnarray}
where the dimensionless frequency $\tw$ is defined by
\begin{equation}
    \tw =  \frac{\omega}{\omega_c}.
\label{eq:tomega}
\end{equation}
One could now perform  an integration by parts in Eq.~(\ref{here}) and assume that at each end the integrands vanish. However as the integrands are oscillatory,  as the term in curly brackets diverges at $|\eta|\rightarrow \infty$, this is potentially dangerous
and requires a choice of regularisation. Instead we proceed by writing the $\cosh(\eta)$ and $\sinh(\eta)$ terms in the curly bracket in Eq.~\eqref{here} in terms of exponentials. Then, as shown in Appendix \ref{app:spectrum}, $\tilde{I}^{k\ell}$ can be written in terms of integrals of the form
\begin{equation}
    \int d\eta \; e^{\nu \eta - q \sinh \eta} = i \pi H_\nu^{(1)}(q) \equiv  i \pi Z_\nu(q)
    \label{eq:H1def}
\end{equation}
where $H_\nu^{(1)}(q)$ are Hankel functions \cite{GR} whose arguments take the explicit values
\begin{eqnarray}
     q &=& i e \tw ,
     \label{eq:qdef}
     \\
     \nu &=& i\tw + s,\qquad {\text{and}} \qquad s=-3\dots 3.
     \label{eq:nuinH}
\end{eqnarray}
The explicit expressions for $\tilde{I}^{k\ell}$ are given in Eqs.~\eqref{eq:Iqq} and \eqref{eq:tI12}, from which $\pgw(\omega)$ is directly obtained by substituting Eq.~\eqref{eq:Pomega}. 
Furthermore, using the recursion relations for Hankel functions one can analytically calculate $\pgw(0)$: as shown in App.~\ref{app:P0}, the result agrees entirely with Eq.~\eqref{eq:P0} thus validating our calculation of $P(\omega)$.

\begin{figure*}[t]\centering
  \includegraphics[width=12cm]
  {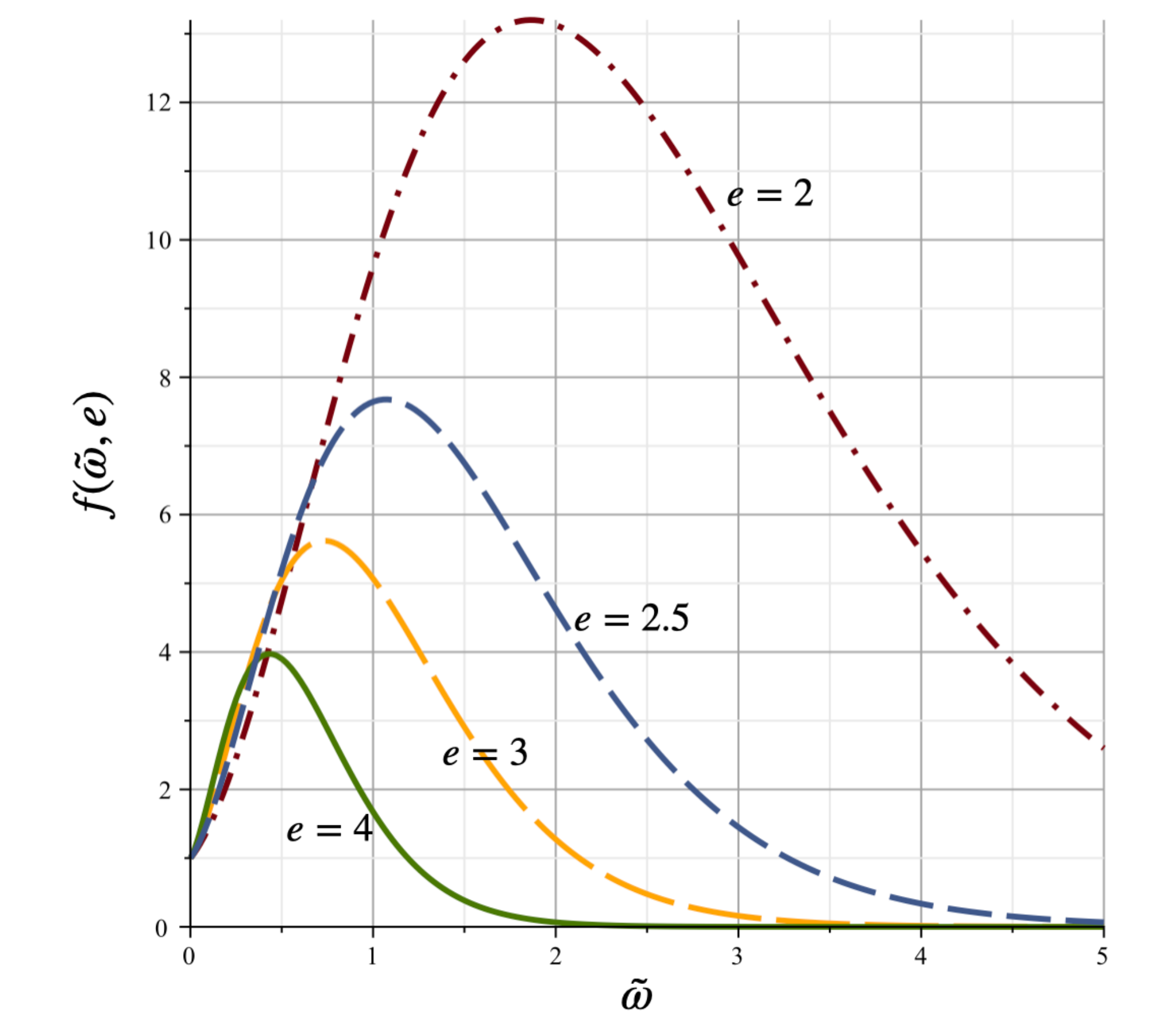}
  \caption{The energy spectrum $f(\tw,e)=\pgw(\tw)/\pgw(0)$ for different eccentricities.}
    \label{fig:dimensionless-power-GR}
\end{figure*}

To summarize, the power emitted in GWs is of the form
\begin{equation}
    \pgw(\omega)=\pgw(0) {f}(\tw,e)
\end{equation}
where $\pgw(0)$ is given in \eqref{eq:P0} and depends on $(p,e,\mu,m)$. The dimensionless ${f}(e,\tw)$ equals 1 at zero frequency; see figure \ref{fig:dimensionless-power-GR} for its general form for different values of $e$.  The maximum GW power is emitted at peak frequency $\tw_{\rm{max}}\sim 6/(e^2-1)$ leading to
\begin{equation}
  \omega_{\rm{max}} \sim 6 \sqrt{\frac{G_Nm(e^2-1)}{p^3}} = 6\sqrt{\frac{G_N m}{r_{\rm min}^3}}\frac{\sqrt{e-1}}{e+1}.
\end{equation}
As expected this scales as $\sqrt{G_N m r_{\rm min}^{-3}}$ with a further $e^{-1/2}$ dependence  for large eccentricity.

\subsection{Kicks}
\label{subsec:kicks}

GWs not only carry away energy but also linear and angular momentum.  The linear momentum radiation leads to a recoil or kick of the CM which can have important astrophysical effects, see e.g.~\cite{Merritt:2004xa}.
Analytical estimates of the recoil velocity started many years ago \cite{Fitchett}, as reviewed in~\cite{Blanchet:2013haa}. For bound BBHs, the understanding that most of the recoil occurs in the strong gravity regime (inside the ISCO through coalesence, merger and ringdown) has led to extensive numerical relativity simulations to study the problem see e.g.~\cite{Baker:2006vn}, as well as approaches using EOB, see e.g.~\cite{Schnittman_2007}. Here we focus on hyperbolic orbits and assume that the analytic calculations have some validity provided the closest distance of approach of the two bodies is sufficiently large that the PN approximation is applicable.  In later sections we will include the effect of scalar radiation on these estimates.

In GR to leading order, the radiation of linear momentum changes to the CM velocity $\vec{V}$ according to
\begin{equation}
    m\frac{dV^i}{dt}=-\cF^i
    \label{eq:CM}
\end{equation}
where the effective force is \cite{poisson_will_2014}
\begin{eqnarray}
    \cF^x&=&F \sin\phi(1+e\cos\phi)^4 \left[1+\frac{175}{58}e\cos\phi +\frac{2}{29}e^2(3+40e\cos^2\phi)+\frac{5}{58}e^3\cos\phi(2+9\cos^2\phi) \right]
    \label{eq:FGRx}
    \\
    \cF^y &=& -F(1+e\cos\phi)^4
    \left[ 
    \cos\phi -\frac{e}{58}(9-175\cos^2\phi)-\frac{e^2}{29}\cos\phi(1-80\cos^2\phi) + \frac{e^3}{58}(2+3\cos^2\phi + 45 \cos^4\phi)
    \right]
    \label{eq:FGRy}
\end{eqnarray}
with
\begin{equation}
    F= F_{\rm GR}\equiv \frac{464}{105}\Delta \eta^2 \frac{c^4}{G_N}\left(\frac{G_Nm}{c^2p}\right)^{11/2}
\end{equation}
and $\Delta$ is the mass difference
\begin{equation}
    \Delta = \frac{m_A-m_B}{m}.
    \label{eq:massdiff}
\end{equation}
For hyperbolic systems, integration of Eq.~\eqref{eq:CM} from $\phi_-$ (with $V^i(\phi_-)=0$) to $\phi$ gives that $V^x(\phi_+)=0$ whereas $V^y(\phi_+) \neq 0$. Thus the CM picks up a velocity in the $y$-direction given by
\begin{equation}
    \Delta V^y_{\rm{GR}} = \frac{5943}{1392} v_{cm}\left[
    \sqrt{1-\frac{1}{e^2}}\left(e^4 + \frac{751}{447} e^2 +\frac{32}{447}\right)
    + \frac{37}{298}e^2 \arccos(-1/e)\left(e^4 + \frac{456}{37}e^2 + \frac{312}{27}\right)
    \right]
    \label{eq:Vygr}
\end{equation}
where
\begin{equation}
    v_{cm} = \frac{464}{105} \Delta \nu^2 \left( \frac{G_Nm}{c^4p}\right)^4 c.
    \label{eq:vcmGR}
\end{equation}
This corresponds to the recoil velocity for a bound system on a circular orbit of radius $p$; for instance if $p\sim 10Gm/c^2$, then $v_{cm}\sim 130 \Delta \nu^2$km/s.

\section{Scalar fields with disformal couplings and precession}
\label{sec:scalar-precession}

In the remainder of this paper our aim  is to extend the results of Sec.~\ref{sec:GReffectsGalore}
to scalar tensor theories \cite{Damour:1992we}. In this section we introduce the theory considered, namely that of a massless scalar field both conformally and disformally coupled to matter \cite{Bekenstein:1992pj}, and then focus on the conservative dynamics deriving the 1PN Einstein Infeld Hoffmann (EIH) Lagrangian in the CM frame, from which we then calculate precession effects.  Linear memory and kicks are considered in later sections.

\subsection{Action and scalar field equation}
\label{subsec:action}

Consider a massless scalar field coupled to matter with dynamics described by the {Einstein frame} action
\be 
S= \int d^4x \sqrt{-g} \left(\frac{1}{16\pi G_N}R-\frac{1}{2}(\partial \phi)^2 \right)+ S_m(\psi,  g^J_{\mu\nu})
\ee
where the matter field $\psi$ couples to the disformal, {Jordan}, metric \cite{Kaloper:2003yf,Brax:2018bow,Kuntz:2019zef,BraxKuntz}
\be 
 g^J_{\mu\nu}\equiv A^2(\phi) g_{\mu\nu} + \frac{2}{\Lambda^2m^2_{\rm Pl}} \partial_\mu \phi \partial_\nu \phi
 \label{eq:jordanmetric}
\ee
where $A(\phi)= e^{\beta \phi/m_{\rm Pl}}\sim 1+ \frac{\beta\phi}{m_{\rm Pl}}$ {is the conformal coupling function} and $\Lambda$ (an inverse length scale) characterises the strength of the disformal interaction.  {These models are particular cases of Horndeski theories \cite{Deffayet:2011gz,Bettoni:2013diz,Sakstein:2015jca,Kobayashi:2019hrl}, and the full transformation between the Einstein frame action to the Jordan frame Horndeski action
can be found in \cite{Bettoni:2013diz}. In particular, the $G_4$ function --- which multiplies the Ricci scalar in the Jordan frame Lagrangian ${\cal L}_J \supset G_4 {m^2_{\rm Pl}}R_J/2$, and which plays a role in the number of gravitational polarisations in the Jordan frame, see below --- reads in our case
\be 
G_4= A^2(\phi)\left(1+ \frac{2(\partial \phi)^2}{A^2(\phi)\Lambda^2m^2_{\rm Pl}}  \right)^{1/2}
\ee
where the contraction is with the Jordan metric (see Appendix C of \cite{Bettoni:2013diz}). This displays a dependence on both the conformal and disformal couplings. } 
 At leading order in an effective expansion where $\phi/m_{\rm Pl} \ll 1 $ and $\partial \ll \Lambda$, the effects of the  conformal and disformal coupling of the scalar field to matter are captured by the scalar field equation 
\be 
\Box \phi= -\frac{\beta}{m_{\rm Pl}} T + \frac{2}{\Lambda^2 m^2_{\rm Pl}}(D_\mu \partial_\nu \phi) T^{\mu\nu}
\label{KG}
\ee
where $D_\mu$ is the covariant derivative of the ambient metric. 
Note that only the energy-momentum tensor of matter, $T^{\mu\nu}$ (with trace $T=T^{\mu}_{\; \; \mu}$), sources the scalar field. At this order matter is conserved 
\begin{equation}
    D_\mu T^{\mu\nu}=0.
    \label{eq:dmuTmunu}
\end{equation} 
The absence of the energy momentum tensor of the scalar on the right hand side of Eq.~(\ref{KG}) means that the scalar field does not source its own propagation equation in a non-linear way, so there is no analogue of the GR non-linear memory effect. Indeed the scalar field equation is linear and can be simply solved iteratively in perturbation theory \cite{BraxDavis} by first considering
\be 
\phi^{(0)}= - \frac{\beta}{m_{\rm Pl}}\Box^{-1}  T
\label{eq:KGscalar}
\ee 
where the retarded Green's function is selected. The disformal interaction is taken into account in a ladder expansion $\phi= \phi^{(0)} + \delta\phi$ where 
\begin{equation}
    \delta \phi= \sum_{n=0}^{\infty} \delta \phi^{(n)}
\end{equation}
with
\begin{eqnarray}
    \Box \delta \phi^{(0)}&=&\frac{2}{\Lambda^2 m^2_{\rm Pl}}(D_\mu \partial_\nu \phi^{(0)}) T^{\mu\nu}
\label{eq:ladder0}
\\
\Box \delta \phi^{(n+1)}&=&\frac{2}{\Lambda^2 m^2_{\rm Pl}}(D_\mu \partial_\nu \delta\phi^{(n)}) T^{\mu\nu} 
\qquad \qquad (n\geq 0).
\label{eq:laddern}
\end{eqnarray}
In the following we will consider the leading  conformal and disformal effects $\phi^{(0)}$, $\delta \phi^{(0)}$.

\subsection{The scalar field emitted by a binary system}

The scalar field sourced by the binary system of point masses $m_{A,B}$ is obtained solving Eqs.~(\ref{eq:KGscalar}) and (\ref{eq:ladder0}), see \cite{BraxDavis}. {Working to order $(v/c)^2$}, $\phi^{(0)}(\vec{x}) = \phi_A^{(0)}(\vec x) + \phi_B^{(0)}(\vec x)$ with the field emitted by body $J=(A,B)$ given by
\be 
\phi_J^{(0)}(\vec x)= -\frac{\beta m_J}{4\pi m_{\rm Pl}}\frac{1- \frac{\vec v_J^2}{2} + \frac{\vec v_{J\perp}^2}{2}}{\vert \vec{x}-\vec{x}_J\vert}
\label{eq:conf}
\ee 
where the perpendicular velocity is orthogonal to the line between body $J$ and the position $\vec x$
\begin{equation}
    \vec{v}_{J\perp} = \vec{v}_J-\vec{n}_J (\vec{v}_J\cdot \vec{n}_J)
    \label{eq:perp}
\end{equation}
and $\vec{n}_J$ is the unit vector between $J$ and the observer: $\vec{n}_J = (\vec{x}-\vec{x}_J)/|(\vec{x}-\vec{x}_J)|$.  Using this solution as a source for the disformal term via Eq.~(\ref{eq:ladder0}) gives\footnote{note we use slightly different notation from \cite{BraxDavis,BraxKuntz}.}
$\delta \phi^{(0)}(\vec x)= \delta \phi^{(0)}_{A}(\vec x)+\delta\phi^{(0)}_{B}(\vec x)$
with 
\begin{eqnarray}
    \delta \phi_{J}^{(0)}(\vec x)&=& - \frac{\beta G_N m_A m_B}{\pi m_{\rm Pl}}\left[ \frac{ \vec a \cdot \vec r + \vec v^2 - 3 (\vec n \cdot \vec v)^2}{\Lambda^2 \vert \vec{x}-\vec{x}_J\vert  r^3}\right]
    \label{eq:disf}
\end{eqnarray}
where we have used the reduced Planck scale $m_{\rm Pl}^2= (8\pi G_N)^{-1}$.

These first order conformal and disformal corrections, \eqref{eq:conf} and \eqref{eq:disf}, can alternatively be obtained in an effective approach. As seen from afar, the binary system can be described  as a source with an energy momentum $T^{\mu\nu}$ characterised by a multipole expansion, e.g. the monopole corresponding to its total mass and its quadrupole play a prominent role.   This effective approach will be generalised in a later section \ref{sec:eff} and simplifies the study of radiation effects for instance. From Eq.~\eqref{eq:KGscalar},
\begin{eqnarray}
     \phi^{(0)}(t,\vec{x}) &=& \frac{\beta}{4\pi m_{\rm pl}} \int d^3 y \frac{T(t-|\vec{x}-\vec{y}|,\vec{y})}{|\vec{x}-\vec{y}|}
    \simeq \frac{\beta}{4\pi m_{\rm pl}R} \int d^3 y {T(t_R,\vec{y})}
       \nonumber
     \\
       &\simeq& \frac{\beta}{4\pi m_{\rm pl}R} \int d^3 y \left[  {T^{0}_{\; \; 0}(t_R,\vec{y}) + T^{i}_{\; \; i}(t_R,\vec{y})} 
     \right]
     \nonumber
\end{eqnarray}
where the retarded time $t_R\equiv t-R$, $|\vec{x}|\equiv R \gg d$ with $d$ the characteristic size of the source, and we work to lowest order in velocity. Then, on using conservation of energy $\partial_\mu T^{\mu \nu} = 0$ and integrating by parts, it follows that for the binary system
\begin{eqnarray}
    \phi^{(0)}(t,\vec{x})   &\simeq & 
\frac{\beta}{4\pi m_{\rm pl}R} \left[ -m + \frac{\ddot{q}^i_{\;\; i}(t_R)}{2}\right]
\end{eqnarray}
where the quadrupole moment is
\begin{equation}
    {q}^{ij}  = \int d^3 y  T^{00}y^i y^j = \mu r^i r^j
\end{equation}
with $\vec{r}$ the vector between $A$ and $B$. 
Now, due to the extra scalar force felt by matter, Newton's law reads $\vec{a}=-\Geff m/r^3 \vec{r}$ (see Eq.~(\ref{eq:eofmfinal}) below), where
\be
\Geff = G_N(1+2\beta^2).
\ee
Thus it follows that
\begin{eqnarray}
     \phi^{(0)}(t,\vec{x})\simeq 
\frac{\beta}{4\pi m_{\rm pl}R}  \left[ -m + \mu\left(v^2(t_R) - \frac{\Geff m}{r(t_R)}\right)\right].
\end{eqnarray}

The next order correction $\delta \phi^{(0)}$ is solution of Eq.~\eqref{eq:ladder0}, and following a similar calculation gives 
\begin{eqnarray}
    \delta \phi^{(0)} = - \frac{2}{\Lambda^2 m_{\rm Pl}^2 4 \pi R} \int d^3 y S
\end{eqnarray}
where the source is\footnote{{ We have removed the divergences at coinciding points as they vanish. Indeed we evaluate $D_\nu\partial_\mu \phi_A^{(0)} T^{\mu\nu}_A$ by point splitting where the source term $T_B^{\mu\nu}$ is a delta function displaced from $x_A$ by a small distance $\epsilon$. The evaluation of $\phi_A^{(0)}$ at this displaced point is proportional to $1/\epsilon$ whose second derivative  vanishes.}}
\begin{equation}
    S = D_\mu \partial_\nu \phi^{(0)} T^{\mu \nu}= D_\mu \partial_\nu \phi^{(0)}_A T^{\mu\nu}_B + D_\mu \partial_\nu \phi^{(0)}_B T^{\mu\nu}_A.
\end{equation}
To lowest order in $v/c$ we only keep the $00$ component leading to
\be 
\int d^3 y S= m_B \partial_0^2 \phi_A^{(0)}(x_B)+ m_A \partial_0^2 \phi_B^{(0)}(x_A)= 2\mu \partial_0^2 \phi^{(0)}(r).
\ee
The spatial integration  over the (nearly) static sources is given by the $00$ component of the energy momentum tensors which is proportional to the masses of the objects. For point-like objects, the integration reduces to the evaluation of the second time derivatives of the fields $\phi_{A,B}^{(0)}$ at the points $x_{A,B}$ respectively. As the fields only depends on $x-x_{A,B}$ respectively, the end result only depends on $r$.   We now use the expression of $\phi^{(0)}$ at  large distance $R\gg d$, so that $\phi^{(0)} \sim r^{-1}$ see \eqref{eq:conf}. 

This is in the spirit of the effective description  approach where the dynamics of the binary system is described in several stages as one probes larger and larger distances. First of all the sources with a finite size are seen from afar as point-like then the binary system of point sources is characterised as a series of multipole encapsulating the its global features, e.g. its monopole or its quadrupole. This is what allows one to integrate out the internal degrees of freedom, i.e. go from finite size objects to point sources and then simply multipole, in order to obtain the radiation field. As we are not dealing with a quantum field theory but a classical field theory, integrating out is equivalent to solving the equations of motion inside the system and using them as a source for the radiation field. We will see how this generalises in the next section.
Then since
\be 
\partial_0^2\left(\frac{1}{r}\right)= -\frac{\vec a \cdot \vec r + \vec v^2 - 3 (\vec n \cdot \vec v)^2}{r^3}
\ee
we find
\begin{equation}
     \delta \phi^{(0)} = - \frac{2 G_N \beta}{\Lambda^2 m_{\rm Pl}  R r^3} \mu m  (\vec a .\vec r + \vec v^2 - 3 (\vec n.\vec v)^2)
     \label{luck}
\end{equation}
namely (\ref{eq:disf}) in the far way limit when $\vert x- x_{A,B}\vert \sim R$.

These solutions for the scalar field allow one to calculate the Fock action of the binary system. Indeed by replacing the explicit solutions to the scalar field equation and integrating over space, one obtains a correction to the dynamics of the binary system described at 1PN by the EIH action \cite{BraxKuntz}.

\subsection{Conservative dynamics in the centre of mass frame}
\label{subsec:EIH}

We now determine the equations of motion to 1PN in the CM frame.
To order $1/c^2$ and truncated at leading order in $G_N/\Lambda^2$, the Fock action reads \cite{Damour:1992we,BraxKuntz,Kuntz:2019zef} \footnote{We have rewritten Eq.~(2.31) of \cite{BraxKuntz}, using the identity $\vec v_{A\perp}. \vec v_{B\perp}= \vec v_A \cdot \vec v_B- (\vec v_A \cdot \vec n) (\vec v_B \cdot\vec n)$ which follows from Eq.~\eqref{eq:perp}.}
\begin{eqnarray}
 {\cal L}_{AB}&=&
\frac{1}{2} m_A \vec v_A^2 +\frac{1}{2} m_B \vec v_B^2 + \frac{\Geff m_Am_B}{r}  +\frac{1}{8c^2} \left[  m_A \vec v_A^4 + {m_B}\vec v_B^4\right] \nonumber \\
& + &\frac{G_N m_A m_B}{2rc^2 }\left[ (3-2\beta^2)( \vec v_A^2+\vec v_B^2) -(7-2\beta^2) \vec v_A \cdot \vec v_B - (1+2\beta^2) \frac{1}{r^2}(\vec{v}_A\cdot \vec{r})(\vec{v}_B\cdot \vec{r})\right]\nonumber\\
 &+& \frac{4\beta^2 G^2_N}{\Lambda^2 r^6}m_Am_B m (\vec r\cdot \vec v)^2-\frac{(\Geff)^2 m_Am_B m}{2 r^2},
\label{eq:action2}
\end{eqnarray}
where as above
\begin{equation}
  \Geff=b\, G_N ,  \qquad    b\equiv 1+ 2\beta^2.
    \label{eq:bdef}
\end{equation}
From the Euler-Lagrange equations there is the conserved quantity
\begin{eqnarray}
    \vec{P} &\equiv & \frac{\partial \cL}{\partial \vec{v}_A} + \frac{\partial \cL}{\partial \vec{v}_B}  
     \nonumber
\\
   &=& m_A \vec{v}_A + m_B \vec{v}_B + \frac{1}{2c^2} \left[ m_A \vec{v}_A v_A^2 + m_B \vec{v}_B v_B^2\right]
    \nonumber
    \\
    &-& \frac{\Geff m_A m_B}{2c^2 r}  \left[ (\vec{v}_A + 
    \vec{v}_B) + \frac{\vec{r}}{r^2}\left( (\vec{v}_A \cdot \vec{r}) + (\vec{v}_B \cdot \vec{r}) \right) \right],
    \label{eq:Pconsv}
\end{eqnarray}
which has no contribution from the disformal factor. Eq.~\eqref{eq:Pconsv} is identical to the standard GR expression (see e.g.~\cite{old-paper}) modulo the expected rescaling of Newtons constant.
One can straightforwardly rewrite $\vec{P}$ as $\vec{P} = \frac{d}{dt} \left[ m_{gA} \vec{r}_A + m_{gB} \vec{r}_B\right]$
where the gravitational masses are given by
\begin{equation}
    m_{gJ} = m_J\left[ 1+ \frac{v_J^2}{2c^2} - \frac{\Geff m_I}{2rc^2}\right] \qquad I \neq J
\end{equation}
and $d[{m_{gA}+m_{gB}}]/dt=0$ to order $1/c^2$.
The system's barycentre is therefore
\begin{equation}
    m\vec{R}_b = m_{gA} \vec{r}_A + m_{gB} \vec{r}_B
\end{equation}
and we now work in the CM frame where $\vec{R}_b=0$. Then the positions of the bodies $\vec{r}_J$ are expressed in terms of $\vec{r}$ as (working to first order in $1/c^2$) 
\begin{eqnarray}
    \vec{r}_A &=& \frac{m_B}{m}\vec{r} + \frac{\nu \Delta}{2c^2} \left(  v^2 - \frac{\Geff m}{r}
\right) \vec{r}
\label{eq:rcmA}
\\
 \vec{r}_B &=& -\frac{m_A}{m}\vec{r} + \frac{\nu \Delta}{2c^2} \left(  v^2 - \frac{\Geff m}{r}
\right) \vec{r}
\label{eq:rcmB}
\end{eqnarray}
where as above $\nu = \frac{m_A m_B}{m^2} = \frac{\mu}{m}$ and $\Delta = (m_A-m_B)/m$, and to leading order
\begin{equation}
    \vec{v}_A = \frac{m_B}{m} \vec{v} , \qquad \vec{v}_B = -\frac{m_A}{m} \vec{v} .
    \label{eq:vcm}
\end{equation}
Finally, substitution of \eqref{eq:rcmA}-\eqref{eq:vcm} into Eq.\eqref{eq:action2} gives the Fock action in the CM frame including the massless scalar field conformal and disformal interactions\footnote{This corrects the expression in \cite{Benisty:2022lox} derived from \cite{Brax:2018bow} where acceleration terms had been dropped.}\footnote{This reduces to the usual EIH GR Lagrangian when $\beta=0=\Lambda^{-2}$.}
\begin{eqnarray}
\frac{ {\cal L}_{0}}{\mu} &=&
\frac{1}{2} \vec v^2 +\frac{\Geff m}{r}  +\frac{(1-3\nu)}{8c^2} (\vec{v}^2)^2 \nonumber \\
& + &\frac{G_N m}{2rc^2 }\left[ \left( (3+\nu) -(1-\nu)2\beta^2\right) \vec{v}^2 + (1+2\beta^2) \frac{\nu}{r^2} (\vec{r}\cdot \vec{v})^2 \right]\nonumber\\
 &+& \frac{4\beta^2 G^2_N}{\Lambda^2 r^6}m^2(\vec r\cdot \vec v)^2-\frac{(1+2\beta^2)}{2c^2}\left(\frac{G_N m }{ r}\right)^2 + {\cal{O}}(c^{-4},\beta^4,\Lambda^{-4}).
\label{eq:actionCM}
\end{eqnarray}
Eq.~\eqref{eq:actionCM}  differs from equation (27) of \cite{BraxBenisty} since there the disformal effects where only accounted in a Post-Minkowskian limit as in \cite{BraxDavis}. The expression we give here is the full 1PN action following \cite{BraxKuntz}.   
The equations of motion in the CM frame follow directly: 
to lowest order, Newton's equations are
\begin{equation}
    \vec{a} = -\frac{\Geff m}{r^3} \vec{r},
    \label{eq:Newtonderived}
\end{equation}
and working to 1PN gives
\begin{eqnarray}
\vec{a} &=& - \frac{\Geff  m}{r^3} \vec{r} - \frac{G_N m}{r^3c^2}\vec{r} \left\{ 
{v^2}C_1 -\frac{\Geff  m}{r}C_2 - \frac{3b\nu}{2r^2}(\vec{r}\cdot \vec{v})^2
\right\}
\nonumber
\\
&+& \frac{2G_Nm}{r^3c^2} \vec{v}
(\vec{r}\cdot \vec{v})
[2 - \nu +2\nu\beta^2]
+
\frac{8\beta^2 G_N^2 m^2  \vec{r}}{\Lambda^2 r^6} \left[ \frac{3(\vec{r}\cdot \vec{v})^2}{r^2} + \frac{\Geff  m}{r} - {v^2}
\right]
\label{eq:eofmfinal}
\end{eqnarray}
where
\begin{eqnarray}
   C_1 &\equiv& 
   1+3\nu + 2\beta^2(3\nu-1)
\nonumber
\\
  C_2 &\equiv&   
  4+2\nu + 4\beta^2\nu.
\nonumber
\end{eqnarray}

\subsection{Secular effects: precession}
\label{sec:prec}

We now determine how the scalar interactions affects the secular evolution of the eccentricity, periastrion and also precession of binary systems.
First we rewrite Eq.~\eqref{eq:eofmfinal} as
\begin{equation}
    \vec{a} = -\frac{\Geff  m}{r^2}\vec{n} + \left[ \bar{R} \vec{n} + \bar{S} \vec{\lambda}\right]
\end{equation}
where recall that $\vec{v}=\dot{r}\vec{n} + (r\dot{\phi})\vec{\lambda}$, and
\begin{eqnarray}
   \bar{R} &=& - \frac{G_N m}{r^2c^2} \left\{ 
{v^2}C_1 -\frac{\Geff  m}{r}C_2 - \dot{r}^2 C_3 \right\}
+  \frac{8 \beta^2 G_N^2 m^2}{\Lambda^2 r^5 }  \left[ 3 \dot{r}^2  + \left(\frac{\Geff  m}{r} - v^2 \right)
\right]
\label{eq:Rgen}
\\
    \bar{S} &=& \frac{G_N m}{rc^2}   C_4 \dot{r}  \dot{\phi}
 \label{eq:Sgen}   
\end{eqnarray}
with
\begin{eqnarray}
    C_3 &=& 4-\frac{\nu}{2}-\beta^2 \nu
    \\
    C_4 &=& 4-2\nu-4\beta^2\nu.
\end{eqnarray}
While conformal effects appear through $\beta$ corrections in the coefficients $C_i$ ($i=1\ldots 4)$, the main effect is the disformal term which changes the component of the force in the direction $\vec{n}$ as appearing in  $\bar{R}$. The component of the force $\bar{S}$ is independent of the disformal term.

Now eccentricity, periastron, and precession evolve according to \cite{poisson_will_2014}
\begin{eqnarray}
 \frac{dp}{df} & \simeq& 2 \frac{p^3}{\Geff  m} \frac{\bar{S}}{(1+e\cos f)^3} 
    \label{eq:pevoln}
    \\
     \frac{de}{df} &\simeq & \frac{p^2}{\Geff  m} \left[  \frac{\sin f}{(1+\cos f)^2} \bar{R} + \frac{2\cos f+e(1+\cos^2 f)}{(1+e\cos f)^3} \bar{S} \right]
     \label{eq:eevoln}
     \\
    \frac{dw}{df} &\simeq& \frac{1}{e} \frac{p^2}{\Geff  m} \left[ - \frac{\cos f}{(1+e\cos f)^2} \bar{R} + \frac{(2+e\cos f)\sin f}{(1+e\cos f)^3} \bar{S} \right]
    \label{eq:periastrionScalar}
\end{eqnarray}
where, as in Sec.~\ref{sec:GReffectsGalore}, $ r={p}/(1+e\cos f)$. These equations can be solved after substituting $\bar{R}$ and $\bar{S}$ from Eqs.~\eqref{eq:Rgen}, \eqref{eq:Sgen}, together with the Newtonian equations, $\dot{r} =q e\sin f,
r\dot{\phi} = q(1+e\cos f)$ with $q=\sqrt{{\Geff  m}/{p}}$.

Concerning the evolution of the periastrion, it follows from Eq.~\eqref{eq:pevoln} that for elliptical orbits ($0\leq f<2\pi$)  there are no secular changes in $p$. The same is true for hyperbolic orbits. 
Similarly, for both for elliptical and hyperbolic orbits, there is no secular evolution of the eccentricity $e$ as a consequence of Eq.~\eqref{eq:eevoln}. As in GR, these results are a consequence of angular momentum and energy and conservation.
Concerning precession $w$, one finds
\begin{eqnarray}
    \frac{dw}{df} &\simeq & -\frac{G_N m}{pec^2} \left\{
    \left(2\beta^2-3\right)e + \left[ 
    (-4C_1 + C_3 - C_4)\frac{e^2}{4} - C_1 + C_2
    \right]\cos(f) \right.
    \nonumber\\
   && \qquad \qquad \left. +  \left(C_4 - C_1 + \frac{C_2}{2}\right) e \cos(2f)+ \frac{e^2}{4}(C_4-C_3)\cos(3f)
    \right\}
     \nonumber\\
   && + \frac{8\beta^2 G_N^2 m^2}{\Lambda^2 p^4} \cos(f)(1+e\cos(f))^3 \left(
   3e\cos^2(f) + \cos(f) -2e
  \right\}.
\end{eqnarray}
For bound orbits, other than the disformal term, only the first term on the first line contributes to the integral leading to
\begin{equation}
      w = \int_0^{2\pi} \frac{dw}{df} df = 2\pi \Delta_p 
\end{equation}
where 
\begin{eqnarray}
     \Delta_p&=&  
\Delta^{\rm{GR}}\left[
     \left(1-\frac{2}{3}\beta^2\right) + \frac{4}{3}\epsilon_\Lambda 
     \left(1 + 3 e^2 + \frac{3}{8}e^4  \right)
     \right]
\end{eqnarray}
with $\Delta^{\rm{GR}}$
the standard GR result given in \eqref{eq:DeltaGR}, and $\epsilon_\Lambda$ the relative magnitude between the disformal and the GR contributions given in the introduction.
The first conformal term agrees with \cite{Will:2014kxa}. Notice, however, that the disformal contribution disagrees with \cite{BraxDavis} since it has a prefactor of unity contrary to a factor five. The difference is accounted for by acceleration terms in the action which were dropped in \cite{BraxDavis}.

For hyperbolic orbits one finds  
\begin{eqnarray}
     w &=& \int_{\phi_-}^{\phi_+} \frac{dw}{df} df 
     \nonumber
     \\
     &=&
{2} \Delta_p \arccos(-1/e) 
      \nonumber
     \\
     &+&
      \sqrt{e^2-1} \frac{\Delta^{\rm GR}}{3e^2} 
      \left\{\Big[ 
     2(2+e^2) + 5\nu(e^2-1) - \beta^2\left(1+\nu(e^2-1)+4e^2\right)
     \Big]+ 
 e^2 \epsilon_\Lambda \left[ 
\frac{5}{3}(11e^2+10) 
\right] \right\}.
\end{eqnarray}
Again there are both conformal and disformal effects, with the latter being determined by the ratio $\epsilon_\Lambda$.

\section{The scalar memory}
\label{sec:scalar-memory}

We now consider the radiative sector and determine the contributions of the scalar field to the linear memory effect.

\subsection{The effective approach}
\label{sec:eff}

Scalar and gravitational radiation emission in conformal coupled theories has been studied in \cite{Vernizzi-Kuntz} and extended to disformal theories in \cite{BraxKuntz}. As shown there, after integrating out the short wavelength modes, the effective action $S_{\rm eff}[x_A,\bar{h}_{ij},\bar{\phi}]$ for the long wavelength radiated fields  can be expanded as
\begin{equation}
    S_{\rm eff}[x_A,\bar{h}_{ij},\bar{\phi}] = S_0[x_A] + S_1[x_A,\bar{h}_{ij},\bar{\phi}]+ S_2[x_A,\bar{h}_{ij},\bar{\phi}]+ S_{\rm NL}[x_A,\bar{h}_{ij},\bar{\phi}]
\end{equation}
where $\bar{h}_{ij}$ is the transverse and traceless metric perturbation, $S_0$ is the conservative action corresponding to the Lagrangian ${\cal L}_{0}$ of \eqref{eq:actionCM}. The terms linear and quadratic in the radiating fields are respectively $S_{1,2}$ whereas $S_{\rm NL}$ contains higher-order non-linear terms than those we consider. Here we follow the standard methods expounded in \cite{Goldberger:2004jt,Porto:2016pyg}.
The gravitational interaction for the binary system takes the standard form 
\begin{equation}
    S_{1}^{(h)} =-\frac{1}{4}\int dt I^{ij} \ddot{\bar{h}}_{ij}
\end{equation}
where $I^{ij}$ is the quadrupole moment of the stress energy tensor given in Eq.~\eqref{eq:Idefreal}.
Similarly the radiative scalar interaction term, including also the quadratic term giving the dynamics of the scalar field, is \cite{Vernizzi-Kuntz,BraxKuntz}
\begin{equation}
S^{(\phi)}_{\rm eff}=-\int d^4x \frac{(\partial \bar\phi)^2}{2}+  \frac{1}{m_{\rm Pl}}\int dt \left[ \bar{\phi} I_\phi + \partial_i \bar{\phi} I^i_\phi + \frac{1}{2} I^{ij}_\phi \partial_i\partial_j \bar{\phi}  +\dots \right]
\label{eq:phiradiativeaction}
\end{equation}
where we have truncated the expansion at second order in the derivatives. 
The scalar monopole, dipole and quadrupule moments are 
obtained by integration over an effective source $J$ as, at leading order, 
\begin{equation}
I_\phi= \int d^3 x \left(J + \frac{1}{6}\ddot J x^2\right),\quad I^i_\phi= \int d^3 x \, x^i J, \quad I_\phi^{ij}=\int d^3x \, Q^{ij}_x J,
\label{eq:scalarmoments}
\end{equation}
where 
\begin{equation}
    Q^{ij}_x = x^i x^j - \frac{1}{3}x^2 \delta^{ij},
    \label{eq:Quaddef}
\end{equation}
and the 
term in $\ddot J$ appears to compensate for the tracelessness of $Q^{ij}$. 
The source term takes the form {$J = J_{\rm con} + J_{\rm dis}$} with conformal part (to order $v^2$) \cite{Vernizzi-Kuntz} 
\begin{eqnarray}
\frac{J_{\rm con}}{\beta}&=& -m_A \delta^{(3)}(x-x_A)- m_B\delta^{(3)}(x-x_B)+  ( m_A v_A^2 \delta^{(3)}(x-x_A)\nonumber \\ &&+ m_B v_B^2 \delta^{(3)}(x-x_B)) - \frac{\Geff  m_A m_B}{r}(\delta^{(3)}(x-x_A)+\delta^{(3)}(x-x_B))\nonumber \\
\end{eqnarray}
and disformal part \cite{BraxKuntz}
\be 
J_{\rm dis}= 4\beta \frac{G_N m_Am_B}{\Lambda^2 }\left(\delta^{(3)}(x-x_B)+\delta^{(3)}(x-x_A)\right) \frac{d^2}{dt^2}\left(\frac{1}{r}\right).
\ee
Working in the CM frame, to leading order in $\Lambda^{-2}$, and assuming that the disformal corrections dominate over the higher order PN correction terms, it follows from 
Eq.~\eqref{eq:scalarmoments} that \cite{BraxKuntz}
\begin{eqnarray}
    I_\phi &=& -\beta m + \frac{8\beta \Geff \mu m }{3r}+ \frac{8 G_N {\beta}\mu m }{\Lambda^2}  \frac{d^2}{dt^2}\left(\frac{1}{r}\right)
    \label{eq:Iphi}
    \\
    I_\phi^i &=&
  - 4 \beta \frac{ G_N \mu m }{\Lambda^2}\Delta r^i  \frac{d^2}{dt^2}\left(\frac{1}{r}\right) 
   \label{eq:Iphii}
   \\
   I^{ij}_\phi &=& -\alpha(r)\beta \mu Q^{ij}
   \label{eq:Iphiij}
\end{eqnarray}
where
\begin{equation}
    Q^{ij} = r^i r^j - \frac{1}{3}r^2 \delta^{ij},
    \label{eq:Qdef}
\end{equation}
and
\begin{equation}
    \alpha(r)=1 +4 \left( 2\nu-1 \right)\frac{G_N  m}{\Lambda^2 } \frac{d^2}{dt^2}\left(\frac{1}{r}\right).
\label{eq:alpha1dedf}
\end{equation}

From Eq.~\eqref{eq:phiradiativeaction} the scalar field equation in the presence of the sources in the radiation regime is now 
\be 
m_{\rm Pl}\Box \bar{\phi}= - I_\phi \delta^{(3)}(x) + \partial_i( I_\phi^i \delta^{(3)}(x))- \partial_{i}\partial_j \left(\frac{I_\phi^{ij} }{2}\delta^{(3)}(x)\right)
\ee
whose solution, using the retarded Green function, is 
\be
\bar{\phi}(\vec x,t)\simeq \frac{1}{4\pi m_{\rm Pl} R}{ \cQ}(t-R),
\label{eq:solnbarphi}
\ee
where $\vec{x}=R\vec{N}$ and the effective charge is
\be
\cQ= I_\phi+ N_i \dot I^i_\phi + N_i N_j \frac{\ddot I^{ij}_\phi }{2}
\label{eq:calQdef}.
\ee
We also have 
\be
\partial_i \bar{\phi} \simeq -\frac{N_i \partial_0 \cQ}{4\pi m_{\rm Pl}R}.
\label{eq:partialphi}
\ee
These will be used below to calculate the memory effects. {In the following it should be understood that all quantities are evaluated at the retarded time $t_R = t-R$.}

\subsection{The Jordan frame metric and the Jacobi equation}

We now determine whether the scalar field can induce a linear memory effect for binaries on hyperbolic trajectories. 
As discussion in Subsection \ref{subsec:action}, matter couples to the Jordan metric which itself depends on the scalar field. To linear order in $h_{\mu\nu}$, $\beta$ and $1/\Lambda^2$,  the gravitational perturbation is given by 
\be 
g_{\mu\nu}^J= \left(1+ \frac{2\beta \bar{\phi}}{m_{\rm Pl}}\right)\eta_{\mu\nu}+ \frac{2}{m_{\rm Pl}^2\Lambda^2}\partial_\mu \bar{\phi} \partial_\nu \bar{\phi} + h_{\mu \nu}.
\label{jordan}
\ee
We are interested in the effects of the scalar field generated by a hyperbolic encounter on rods placed at a distance $R$. The field due to the binary system is $\bar \phi$ defined in (\ref{eq:calQdef}).

The effects of a wave on a detector is determined by the variation of its length due to the passage of the wave. This is determined by using the Jacobi equation for geodesic deviations. We will assume that the detector is non-relativistic and denote by $\xi^i$ the vector determining its length. As matter is coupled to the Jordan metric, the Jacobi equation is given by
\be 
\frac{d^2 \xi^i}{dt^2}=-R^i_{0j0}\xi^j
\ee
where the linearised Riemann tensor can be calculated directly from \eqref{jordan} in the Jordan frame.
Choosing coordinates such that the metric perturbation $h_{\mu\nu}$ is in the TT gauge and using the invariance of the linearised Riemann tensor to go to the detector frame where the Jacobi equation is valid, we can use (\ref{jordan}) to find
\be 
R_{i0j0}=\frac{1}{m_{\rm Pl}^2\Lambda^2}\left[2(\partial_i\partial_j\bar{\phi}) (\partial_0^2 \bar{\phi}) + \partial_j\bar{\phi} \partial_i\partial_0^2\bar{\phi} + \partial_i\bar{\phi} \partial_j\partial_0^2 \bar{\phi}\right] +\frac{\beta}{m_{\rm Pl}}\partial_i\partial_j \bar{\phi} -\frac{1}{2} \partial^2_0 h_{ij}^J
\label{eq:Rphil}
\ee
where, similarly to GR, we have made explicit the dependence on the Jordan metric perturbation comprising both the scalar and gravitational effects  
\begin{equation}
    h_{ij}^J =  \frac{2\beta \bar{\phi}}{m_{\rm Pl}}\delta_{ij}+ \frac{2}{m_{\rm Pl}^2\Lambda^2}\partial_i \bar{\phi} \partial_j \bar{\phi} + h_{ij}^{TT}.
    \label{eq:hjordan}
\end{equation}
Notice that the disformal part contributes to non-linear order. At linear order, the contribution from the conformal part reads
\be
R_{0i0j} = -\frac{1}{2}\partial_0^2 A_{ij}
\ee
where
\be
A_{ij} 
\supset \frac{2\beta}{m_{\rm Pl}} \bar{\phi} (\delta_{ij}- N_i N_j).
\ee
The longitudinal part vanishes at linear order, leaving only a transverse component. This agrees entirely with the results of \cite{Heisenberg:2024cjk} who find $A_{ij} 
\supset P_b (\delta_{ij}- N_i N_j)$ where $P_b = \sigma \bar \phi$ with
\be
\sigma = \left. \frac{G_{4,\phi}}{{G_4}} \right|_{\phi = \phi_0}=
\frac{2\beta}{m_{\rm Pl}} \left(1- \frac{(\partial \phi_0)^2}{m_{\rm Pl}^2 \Lambda^2}\right) \sim \frac{2\beta}{m_{\rm Pl}} .
\ee

Defining $\xi^i= \xi^i_0 +\delta \xi^i$ where $\xi^i_0$ is the position of the detector prior to the passage of any gravitational wave or scalar, we thus have to linear order in perturbations $\delta \xi^{I}$
\be 
\frac{d^2 \delta \xi^i}{dt^2}= -\left\{ \frac{1}{m_{\rm Pl}^2\Lambda^2}\left[2(\partial^i\partial_j \bar\phi) (\partial_0^2 \bar\phi) + \partial_j\bar\phi \partial^i\partial_0^2\bar\phi + \partial^i\bar\phi \partial_j\partial_0^2 \bar\phi\right] +\frac{\beta}{m_{\rm Pl}}\partial^i\partial_j \bar\phi -\frac{1}{2} \partial^2_0 (h^{i}_{\;  j})^J\right\} \xi^j_0
\ee
with solution
\be 
\delta \xi^i= \delta \xi^i_J+ \delta \xi^i_L + \delta \xi^i_{NL}
\ee
where
the Jordan equivalent to the GR case is
\be 
\delta \xi^i_J= \frac{1}{2}  h_{ij}^J\xi^j_0.
\ee
There is  a {\it linear scalar memory effect} (denoted by $L$)
\be 
\delta \xi^i_L= -\frac{\beta}{m_{\rm Pl}}\partial_i\partial_j\left( \int_{-\infty}^t d\tau \int_{-\infty}^\tau d\tau'\bar\phi (x,\tau')\right) \xi^j_0
\label{lin}
\ee
which depends on the conformal coupling $\beta$ as well as a {\it non-linear scalar memory effect} (denoted by NL) depending quadratically on the history of the scalar field
\be 
\delta \xi^i_{NL}= -\int_{-\infty}^t d\tau \int_{-\infty}^\tau d\tau'\frac{1}{m_{\rm Pl}^2\Lambda^2}\left[2(\partial_i\partial_j \bar\phi) (\partial_0^2 \bar\phi) + \partial_j\bar\phi \partial_i\partial_0^2\bar\phi + \partial_i\bar\phi \partial_j\partial_0^2 \bar\phi\right] (x,\tau') \xi^j_0.
\label{nl}
\ee
These are particular to the presence of couplings between matter and the scalar field. Notice that this non-linear effect comes from the disformal coupling which is quadratic in the scalar field and differs from a potential non-linear effect of the scalar in the scalar field equation. As already mentioned, only the matter energy-momentum tensor acts as a source for $\phi$ and there is no non-linear scalar effect  per se in the scalar field equation\footnote{Non-linear effects in the scalar field equation would appear if the scalar has self-interactions which are not taken into account here.}.

For a rod defined by $\xi^i$, the variation of the Euclidean length is 
\be 
\delta L= \frac{\delta \xi^i \xi_i^0}{L}
\ee
where $L= \sqrt{\xi^0_i \xi_0^i} $, implying that
\be 
\frac{\delta L}{L}= \frac{\delta \xi^i}{L} \hat{\ell}_i =  \left. \frac{\delta L}{L}\right|_{J}+ \left.\frac{\delta L}{L}\right|_L +  \left. \frac{\delta L}{L}\right|_{NL}
\ee
where $\hat{\ell}^i= \frac{\xi^i_0}{L}$ is the unit vector along the rod, and
\begin{eqnarray}
    &&\left. \frac{\delta L}{L}\right|_J =
    \frac{1}{2} h_{ij}^{TT} \hat{\ell}^i\hat{\ell}^j+\frac{\beta \bar\phi}{m_{\rm Pl}} + \frac{1}{m_{\rm Pl}^2\Lambda^2} (\partial_i \bar\phi)(\partial_j \bar\phi)\hat{\ell}^i \hat{\ell}^j,
    \label{eq:deltalJ} \\
    && \left. \frac{\delta L}{L}\right|_L = -
     \frac{\beta}{ m_{\rm Pl}}\hat{\ell}^i \hat{\ell}^j \partial_i\partial_j\left( \int_{-\infty}^t d\tau \int_{-\infty}^\tau d\tau'\bar\phi (x,\tau')\right),
      \label{eq:deltalL}  \\
     &&\left.\frac{\delta L}{L}\right|_{NL} = - \hat{\ell}^i \hat{\ell}^j \int_{-\infty}^t d\tau \int_{-\infty}^\tau d\tau'\frac{1}{m_{\rm Pl}^2\Lambda^2}\left[2(\partial_i\partial_j \bar\phi) (\partial_0^2 \bar\phi) + \partial_j\bar\phi \partial_i\partial_0^2\bar\phi + \partial_i\bar\phi \partial_j\partial_0^2 \bar\phi\right] (x,\tau') 
      \label{eq:deltalNL}
 \end{eqnarray}
Of course in GR only the first term in the Jordan  part is non-vanishing.
We will evaluate these different terms when an observer sees the hyperbolic encounter between two massive objects and consider the asymptotic effects between the initial deviation at $t=-\infty$ and the final one at $t=+\infty$. We define
\be
\delta_A = \left. \frac{\delta L}{L}\right|_A (+\infty)-\left. \frac{\delta L}{L}\right|_A (-\infty), \qquad A = J,L,NL
\ee

\subsection{Scalar memory effects}

We now calculate these different terms.
The simplest is the Jordan displacement $\delta_J$.
{From \eqref{eq:deltalJ}, its GR contribution is exactly that given in Eq.~\eqref{eq:delta2GR}.}
The contribution from the radiated scalar follows from \eqref{eq:hjordan} which at leading order reads 
\be 
h^\phi_{ij}= \frac{2\beta}{m_{\rm Pl}}\bar\phi^{}(\vec x)\delta_{ij} + \frac{2}{\Lambda^2 m_{\rm Pl}^2 }\partial_i \bar\phi^{}(\vec x) \partial_i \bar\phi^{}(\vec x) =  \frac{4\beta G_N  \cQ}{R}\delta_{ij} + \frac{8 G_N^2 \dot \cQ^2}{\Lambda^2 R^2 }N_i N_j
\label{asym}
\ee
where all the quantities are evaluated at the retarded time. 
This is a new contribution which complements the gravitational part. Notice that the conformal part is a dilation effect proportional to $\delta_{ij}$. In the case of massive gravitation, the coupling would be $\beta=1/\sqrt 6$. When solar system tests are considered, the Cassini bound leads to a small contribution as $\beta^2 \lesssim 2.1 \cdot 10^{-5}$. On the other hand in the vicinity of compact objects, scalarisation could take place with $\beta={\cal O}(1)$. The disformal part introduces a longitudinal polarisation which is orthogonal to the usual gravitational effect.   
Substituting into \eqref{eq:deltalJ} leads to 
\be 
\delta_J^{(\phi)}= \frac{2\beta G_N }{R}\left[ \cQ(+\infty) -\cQ(-\infty)\right]
+\frac{4 G_N^2 }{\Lambda^2 R^2 }(\vec{N}\cdot\hat{\ell})^2\left[
\dot \cQ^2(+\infty)-\dot \cQ^2(-\infty)\right].
\ee

The integrated memory effects $\delta_L$ and $\delta_{NL}$ (see \eqref{lin} and \eqref{nl}) 
can be expressed as 
\be 
\delta_f = \int_{-\infty}^{+\infty} dt \int_{-\infty }^t dt' f(t')= \int_{\mathbb{R}}dt \ (\theta \star f)(t)
\ee
for given functions $f(t)$ and we define $\star$ as the convolution operator. This is defined up to a constant which will be adjusted to remove potential divergences in the integrals.
It is convenient to Fourier transform the previous expression and using the Fourier transform of the Heaviside distribution $\theta$, i.e.~the Sokhotski-Plemelj result $1/i(\omega+i\epsilon)= -i {\cal P}(1/\omega)+\pi \delta(\omega)$ where $\epsilon\to 0$ and ${\cal P}$ is the Cauchy principal value ,  we find that
\be 
\delta_f= -i \lim_{\omega \to 0} \lim_{\epsilon \to 0}  \frac{f(\omega)}{ \omega +i\epsilon }
\ee
where the limit must be renormalised when it diverges for $\omega=0$ corresponding to an IR divergence at $t=\infty$. This is done by subtracting
so that
\be 
\delta_f= -i \lim_{\omega \to 0} \lim_{\epsilon \to 0} \frac{f(\omega)-f(0)}{ \omega+i\epsilon}= \left. -i\frac{df(\omega)}{d\omega}\right|_{\omega=0}.
\ee
In terms of integrals, this reduces to
$ 
\delta_f= -\int_{-\infty }^{+\infty} t f(t) dt.
$
As an example, consider $f=\ddot g$, from which  $f(\omega)= i \omega (\dot g)(\omega)$ as a function of the Fourier transform of the first derivative $\dot g(t)$. Thus
\be 
\delta_f= (\dot g)(\omega)\vert_{\omega=0}= \int_{-\infty}^{+\infty} \dot g(t) dt = g(+\infty)- g(-\infty).
\ee

Let us apply this result first to the {\it linear case} \eqref{eq:deltalL} for which, to leading order
\be
f(t)= -\frac{\beta}{m_{\rm Pl}}\hat{\ell}^i \hat{\ell}^j  \partial_i \partial_j \phi
= -2 G_N \beta (\vec{N}\cdot\hat{\ell})^2  \frac{\partial_0^2 \cQ }{R}
\ee
This gives
\be 
\delta_L= - 2G_N \beta \frac{ (\vec{N}\cdot\hat{\ell})^2}{R}
\left[\cQ(+\infty)-\cQ(-\infty)\right]
\ee
In the {\it non-linear case} we have 
\be 
f(t)= - \frac{ 4G_N^2 }{\Lambda^2}(\vec{N}\cdot\hat{\ell})^2 \frac{ \partial_0^2 (\dot \cQ^2)}{R}
\ee
giving
\be 
\delta_{NL}= -\frac{4 G_N^2 }{R^2\Lambda^2}(\vec{N}\cdot\hat{\ell})^2   \left[\dot \cQ^2 (+\infty)-\dot \cQ^2(-\infty)\right]
\ee
The complete memory effect from the scalar is obtained by adding the three contributions, namely
$\delta = \delta_J^{(
\phi)} + \delta_L + \delta_{NL}$.  The disformal terms completely cancel and we are left with
\be 
\delta= 2G_N \beta \frac{ (1-(\vec{N}\cdot\hat{\ell})^2)}{R}\left[\cQ(+\infty)-\cQ(-\infty)\right]
\label{eq:canc}
\ee
where $\cQ$ has a contribution from the monopole (M), dipole (D) and quadrupole (Q), see \eqref{eq:calQdef}.

In practice the only parts coming from the monopole which do not vanish at large distance are the mass terms. The are time-independent and do not vary between the incoming and outgoing waves. As the dipole falls off as $r$ goes to infinity,  the change of the charge ${\cal Q}$ only depends on the quadrupole
\be 
\delta= -G_N\mu  \beta^2 \frac{ (1-(\vec{N}\cdot\hat{\ell})^2)}{R}\left[ (N_i N_j \ddot Q^{ij})_{+\infty}-(N_i N_j\ddot Q^{ij})_{-\infty} \right].
\ee
This has the same type of structure as in GR up to the geometrical factors and the coupling $\beta^2$, see \eqref{eq:delta2GR}.

We now evaluate this for the binary system on hyperbolic orbits. Using the results of section \ref{sec:GReffectsGalore}, one has
\begin{equation}
    \delta^{(Q)}_\phi = -8 \left(1-(\vec{N}\cdot \hat\ell)^2\right) (N_x N_y)\frac{G_N\mu \beta^2}{R}
\left(\frac{v_\infty^2}{e^2} \right) 
\sqrt{e^2-1} .
\end{equation}
We have emphasized the dependence on the quadrupole only by replacing $\delta \to \delta_\phi^{(Q)}$. 
Interestingly enough, the order of magnitude of the scalar case is the one of GR modulo the term in $\beta^2$ which reduces its effect. 
In particular, the effect can be relevant  for unscreened objects such as neutron stars where a phenomenon such as scalarisation takes place for which $\beta={\cal O}(1)$ for a neutron star whilst $\beta \ll 1$ in the solar system.

\subsection{Emitted Power}

In GR, as we showed in section \ref{subsec:memory}, there is a direct link between the emitted GW power at zero frequency and the memory effect. We now analyse what happens for disformally coupled scalars. 

The scalar power emitted by the source is given by
\be 
P_\phi = \int d^2 S  \vert x\vert^2 N^i \partial_i \bar{\phi} \partial_0 \bar{\phi}= - \frac{1}{16\pi^2 m^2_{\rm Pl}}\int d^2 S (\partial_0 {\cal Q})^2,
\ee
where the second equality uses \eqref{eq:partialphi}.
Substituting $\cQ$ from \eqref{eq:calQdef} it follows from the identities
\be 
\frac{1}{4\pi}\int d^2 S N_i N_j= \frac{1}{3} \delta_{ij}, \ \ \frac{1}{4\pi}\int d^2 S N_i N_j N_k N_l= \frac{1}{15}( \delta_{ij}\delta_{kl}+ \delta_{ik}\delta_{jl}+\delta_{il}\delta_{jk})
\ee
that \cite{Ross:2012fc}
\be 
P_\phi= -2 G_N \left[ \dot I_\phi^2 + \frac{1}{3} (\ddot I_\phi^i)^2 + \frac{1}{30}(\dddot I_\phi^{ij})^2 \right]
\label{eq:powerscalar}
\ee
as already obtained using unitarity in QFT for instance \cite{Kuntz:2019zef}. 
The scalar radiation is therefore also of three types, monopole, dipole and quadrupole. It is clear from Eq.~(\ref{eq:Iphii}) that in the CM frame, $\ddot{I}^i_\phi$ scales as $\Lambda^{-2}$, thus the dipole appears in Eq.~(\ref{eq:powerscalar}) at the $1/\Lambda^4$ level, beyond our approximation and should be neglected. 
Since
\be 
\dddot I_\phi^{ij}= -\beta\mu \left(\alpha(r) \dddot Q^{ij}+ 3 \dot \alpha(r) \ddot Q^{ij} + 3 \ddot \alpha(r) \dot Q^{ij}+ \dddot \alpha(r) Q^{ij}\right)
\ee
with $\alpha(r)$ given above Eq.~\eqref{eq:alpha1dedf}, this implies that the disformal interaction leads to contributions to the emitted power involving the monopole and the first 3 derivatives of the quadrupole at the $1/\Lambda^2$ level.

\begin{figure*}[t]\centering
  \includegraphics[width=12cm]
  {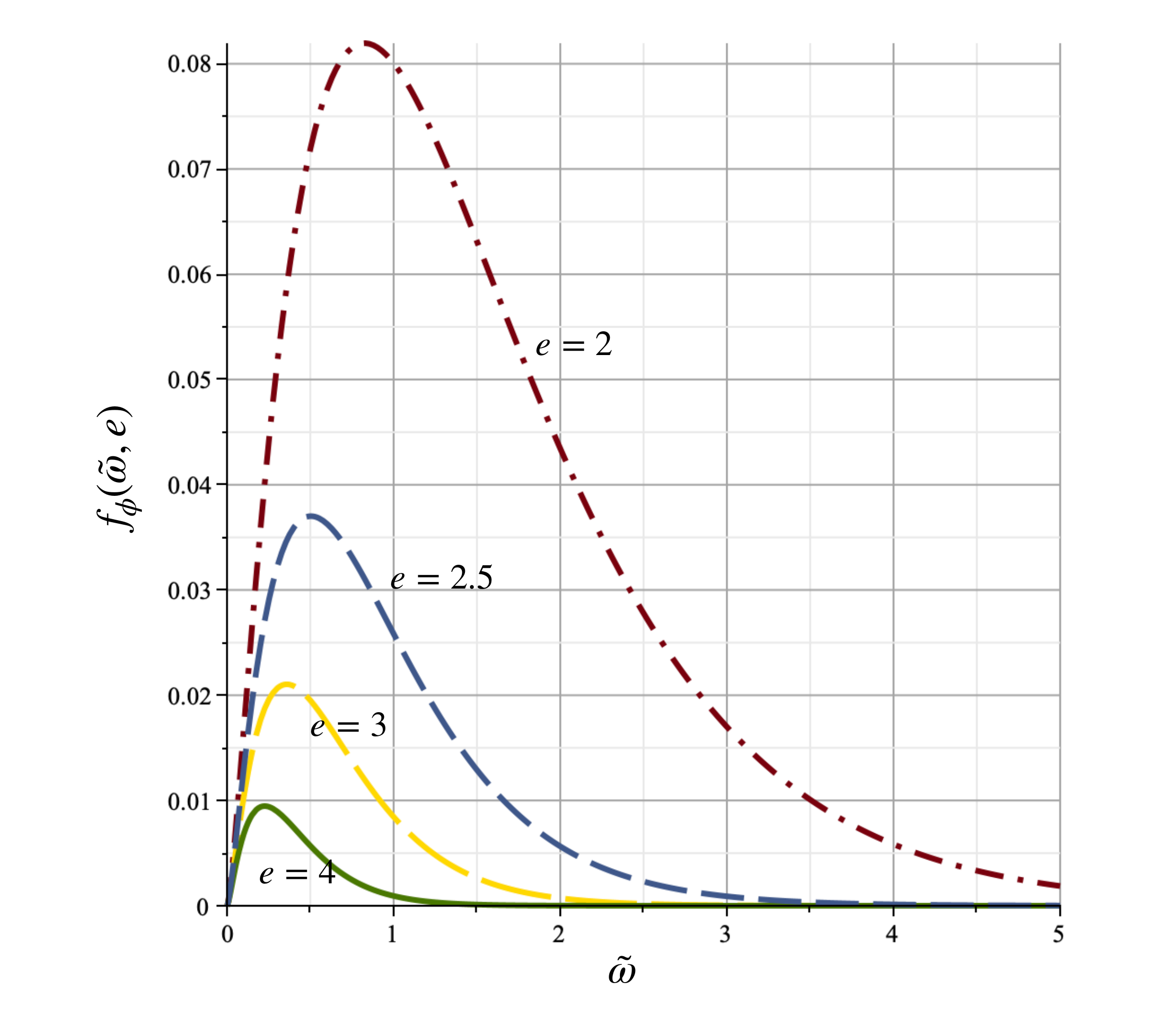}
  \caption{The scalar energy spectrum of the monopole $f_\phi(\tw,e)$ given in \eqref{eq:Phimonopole} for different eccentricities.}
    \label{fig:dimensionless-scalar-power-phi}
\end{figure*}

The monopole contribution depends on the conformal and disformal couplings.
The emitted energy  is given by 
\be 
E_\phi^{(M)}=2 G_N \int dt (\dot I_\phi)^2(t)
=2 G_N\int \slashed{d} \omega \omega ^2 \vert \tilde{I}_\phi (\omega)\vert^2,
\ee
with corresponding power spectrum 
\be 
{\cal{P}}^{(M)}_\phi(\omega)= 2G_N \omega^2 \vert \tilde{I}_\phi (\omega)\vert^2
\ee
where $I_\phi(t)$ is given in Eq.~(\ref{eq:Iphi}).
Note that the constant term $-\beta m$ has a vanishing derivative and hence does not contribute.
We find, using the same approach as in Sec.~\ref{sec:scalar-memory} (working in terms of the mean anomaly) that
\be 
\tilde{I}_\phi(\omega)={\cal X }\frac{e}{a \omega_c}\int_{-\infty}^\infty d\eta e^{-i \tw(e \sinh \eta -\eta)}= {\cal X}\frac{i \pi  e}{a \omega_c} Z_{i\tw}(i\tw e)
\ee
where
\begin{eqnarray}
    {\cal X}&=& 8\beta  \mu m\left( \frac{\Geff}{3}- \frac{\omega^2 G_N}{\Lambda^2}\right) = 8\beta  G_N\mu m\left( \frac{(1+2\beta^2)}{3}- \frac{\omega^2}{\Lambda^2}\right),
\label{Xdef}
\end{eqnarray}
and the correction given by the disformal term cannot exceed unity, i.e.~$\omega \lesssim \Lambda$, as $\Lambda$ is the cut-off scale above which the higher order contributions to the disformal term cannot be controlled. 
In the following we assume that $\omega \ll \Lambda$.  Then
\begin{eqnarray}
{\cal{P}}^{(M)}_\phi(\omega) &=&
\left\{ 2G_N {\cal{X}}^2 \left(\frac{\pi  e}{a}\right)^2
\right\} \left[ \tilde{\omega}^2|Z_{i\tw}(i\tw e)|^2 \right]  \equiv \bar{{\cal P}}_\phi^{(M)} f_\phi(\tw,e).
\label{eq:Phimonopole}
\end{eqnarray}
The $\omega$-independent normalisation is given by
\be
\bar{{\cal P}}_\phi^{(M)} 
\simeq 2G_N \left(\frac{8\beta G_N \mu m}{3}\right)^2 \left(\frac{\pi  e}{a}\right)^2
\ee
assuming $\omega \ll \Lambda$. The frequency dependent part $f_\phi^{(0)}(\tw,e) =\tilde{\omega}^2|Z_{i\tw}(i\tw e)|^2$ vanishes as $\tw\rightarrow 0$
from \eqref{eq:limitZ}, and is shown in figure \ref{fig:dimensionless-scalar-power-phi}, whose functional form should be compared to the GR one in figure \ref{fig:dimensionless-power-GR}. 
The power emitted in the scalar monopole is suppressed relative to the GW power by a factor of
\begin{equation}
    \frac{\bar{\cal P}_\phi^{(M)}}{\pgw(0)} = \left(\frac{8}{3}\right)^2 \frac{5\pi^3}{16} \beta^2 \frac{e^6}{e+1} \frac{G_Nm}{r_{\rm min}},
\end{equation}
namely by the $\beta^2$ term and also $\frac{G_Nm}{r_{\rm min}} \leq 0.5$. On the other hand  it increases with the fifth power of eccentricity.

The quadrupolar term is given by
\be 
E_\phi^{(Q)}=\frac{G_N}{15}\int \slashed{d} \omega \omega^6 \vert \tilde{I}^{ij}_\phi(\omega)\vert^2 
\ee
where
\be 
\tilde{I}^{ij}_\phi (\omega) = -\beta \mu \int dt \alpha(r) (t)  Q^{ij}(t) e^{-i\omega t}.
\label{eq:quadpowerphi}
\ee
{This contribution contains two parts.  The conformal one is proportional to the GW one in GR
\be 
{\cal{P}}_{\phi,\rm{conformal}}^{(Q)}(w)= \frac{\beta^2}{3} \pgw(w)
\ee
and hence does not vanish at zero frequency. 
The disformal part depends on the integral
\be 
u^{ij}(\omega)=\int dt \frac{d^2}{dt^2}\left(\frac{1}{r}\right) Q^{ij}(t) e^{-i\omega t}= \int d\eta f^{ij}(\eta) e^{-i\tilde{\omega}(e\sinh\eta -\eta) }
\ee
where $\tilde{\omega}={\omega}/{\omega_c}$, the tensor $f^{ij}(\eta) \rightarrow $ constant as $|\eta| \rightarrow \infty$.
This implies that $u^{ij}(0)$ diverges and is sensitive to the long time tail of the integrand. As a  result $u^{ij}(\omega)\sim f^{ij}(\infty) \int d\eta e^{-i\tilde{\omega}(e\sinh\eta -\eta)}$. The integral is nothing but a Hankel function 
$i\pi 
Z_{i\tilde{\omega}}(ie \tilde{\omega})$
and  scales like $\propto f^{ij}(\infty) \ln \omega$ when $\omega\to 0$ (using (\ref{eq:limitZ})). As a result $\omega^3 u^{ij} (\omega)$ converges to zero when $\omega \to 0$ and the disformal coupling does not contribute to the zero frequency limit of the quadrupole emission. 
}

Now we have seen that the scalar displacement is proportional to the GR displacement 
\be 
\delta_\phi^{(Q)} \propto \beta^2 \delta^{(2)}_{\rm GR}
\ee
where $\delta^{(2)}_{\rm GR}$ is given in \eqref{eq:delta2GR}. From the above discussion we have 
\be
{\cal{P}}^{(Q)}_\phi(0)= \frac{\beta^2}{3}\pgw(0)
\ee
implying that the quadrupole power spectrum of the scalar field at zero frequency does not vanish. Hence, we have shown that as in the GR case,  the scalar memory effect is determined by  the scalar power spectrum at zero frequency. This results follows from the fact that the monopole power spectrum vanishes and only the quadrupole emission matters. The disformal interaction does not contribute to the memory effect. This is corroborated by the absence of contribution to the power spectrum at zero frequency from the disformal interactions. On the other hand, the detailed relation between the power spectrum at zero frequency and the memory effect is modified by the presence of a scalar field, i.e. if the two  could be extracted from data, their comparison may give a way of looking for the presence of a scalar field conformally coupled to matter and extract its coupling to matter $\beta$.

In the following section, we will find that the disformal interactions do contribute to the kicks on hyperbolic orbits due  to the scalar radiation. This would help disentangling conformal from disformal interactions, of course,  if these effects could be measured with enough precision. 

\section{Scalar kicks}
\label{sec:scalar-kicks}
\subsection{The kicks}
In this last section we determine how the radiation of scalar momentum will modify the GR kick velocity calculated in subsection \ref{subsec:kicks} for hyperbolic orbits. The momentum loss due to scalar emission is given by
\be 
\frac{dP^i}{dt}\supset  - \int d^2S \vert x \vert^2 N_j T^{ij}_\phi 
\ee
where, using Eqs.~(\ref{eq:solnbarphi}), 
\be 
T^{ij}_\phi = \partial^i\bar{\phi} \partial^j \bar{\phi} -\frac{\delta^{ij}}{2}\left(\partial^k \phi \partial_k \bar{\phi} - (\partial_0\bar{\phi})^2\right)
= N^i N^j \frac{(\partial_0 \cQ)^2}{16 \pi^2 m^2_{\rm Pl}\vert \vec{x}\vert^2}
\ee
leading to 
\be 
\frac{dP^i}{dt}\supset - \frac{1}{16\pi^2 m^2_{\rm Pl}} \int d^2S N^i  (\partial_0 \cQ)^2,
\ee
Using the definition of ${\cal Q}$ in Eq.~(\ref{eq:calQdef}), it follows that the only non-vanishing components of this integral involve the second derivative of the scalar dipole moment, namely tensor and scalar contributions add up to
\begin{eqnarray}
    \frac{dP^i}{dt}&=&- \frac{4 G_N}{3}    \dot I_{\phi}\ddot I^{i}_\phi- \frac{8G_N}{15}\ddot I_{j\phi}\dddot I^{ij}_\phi {-} \cF^i.
    \label{eq:momentumradiated}
\end{eqnarray}
{The tensor force is modified from the GR one in Eqs.~\eqref{eq:FGRx}-\eqref{eq:FGRy} as a consequence of the rescaling of Newton's constant in the 0th order equations of motion \eqref{eq:eofmfinal}.}
Thus it is given by Eqs.~\eqref{eq:FGRx}-\eqref{eq:FGRy} {with one main difference: now $F=F_{\rm eff}$ with} 
\begin{equation}
    F_{\rm eff} \equiv F_{\rm GR} \left(\frac{\Geff}{G}\right)^{7/2} = F_{\rm GR}(1+2\beta^2)^{7/2}.
\end{equation}
The calculation of the scalar monopole-dipole and dipole-quadrupole contributions, respectively defined by
\begin{equation}
    \left( {\cal{F}}_\phi^{m-d} \right)^i \equiv \frac{4G_N}{3}  \dot I_{\phi} \ddot I^{i}_\phi \, \qquad  \left({\cal{F}}_\phi^{d-q}\right)^i\equiv \frac{8G_N}{15} \ddot I_{j\phi}\dddot I^{ij}_\phi,
\end{equation}
is detailed in Appendix \ref{app:derivr}. Their exact (and rather long) expressions is given in equations \eqref{eq:annecy1}
 and \eqref{eq:annecy2} respectively, from which one can determine
\begin{equation}
\vec{{\cal{F}}}_\phi =  \vec{{\cal{F}}}_\phi^{m-d} +\vec{{\cal{F}}}_\phi^{d-q} ,
\end{equation}
see \eqref{eq:annecy3}.
Then, the kick is obtained by solving Eq.~(\ref{eq:CM}) namely 
\begin{equation}
    m{\vec{V}}=-\int_{\phi_-}^{\phi_+} d\phi \,\frac{1}{\dot{\phi}} (\vec{\cF}_{\rm{GR}}  +  \vec{{\cal{F}}}_\phi).
    \label{eq:CMbis}
\end{equation}
As in the GR case (see section \ref{subsec:kicks}), the scalar force gives no kick in the $x$ direction. In the $y$-direction, the kick {\it adds} to the GR one $\Delta V^y_{GR}$ of \eqref{eq:Vygr}, and is given by
\begin{equation}
    \Delta V^y_\phi = 
   v_{cm} \epsilon_\Lambda
f_\phi^y(e)
\end{equation}
  where $v_{cm}$ is given in  \eqref{eq:vcmGR}. The function $f_\phi^y(e)$ in given in \eqref{eq:fphiy}, and it should be noted that it scales with larger powers of $e$ (namely $e^8$ rather then $e^4$ in GR, see \eqref{eq:Vygr}). Hence, despite the small coefficient $\epsilon_\Lambda$ in front, these terms could become significant for large eccentricities.
Finally and contrary to the memory effect, the scalar kicks depend crucially on the disformal interaction.

\subsection{Orders of magnitude}

The relative magnitude of the kick due to the disformal interaction compared to the GR case is proportional to $
\epsilon_\Lambda=  \frac{\beta^2 G_N m}{ \Lambda^2 p^3}
$
as mentioned in the introduction. 
This ratio is the one which also governs the magnitude of the precession effects, section \ref{sec:prec}. 
As the disformal coupling that we have considered is the lowest order in a $\partial/\Lambda$ expansion,  the lowest distance of approach has to satisfy
$
p\gtrsim \Lambda^{-1}.
$
As a result, the ratio between the velocity kicks is bounded by
\be 
\epsilon_\Lambda\lesssim \beta^2 G_N m \Lambda.
\ee
This can be reexpressed in terms of the typical size of the objects in the binary system ${\cal R}$ and their Newtonian potential $\Phi_N= \frac{G_N m}{\cal R}$ as
\be 
\epsilon_\Lambda\lesssim \beta^2 \Phi_N {\cal R}\Lambda.
\ee
Typically, we will be interested in White Dwarfs with Newtonian potential of order  $\Phi_N\simeq 10^{-4}$, implying that the disformal kicks can be larger than the GR ones when 
\be 
\Lambda \gtrsim \frac{10^4}{\beta^2 R}\simeq \frac{10^{-10}}{\beta^2}\rm eV
\ee
where we have taken the radius of White Dwarfs to be of the order of $10^4$ km. Taking $\beta={\cal O}(1)$ as the conformal coupling is only constrained by solar system observations and could be larger  for denser compact objects \cite{Ramazanoglu:2016kul}. The distance of closest approach would then  be around 10 km. This should be compared to previous bounds on $\Lambda$ from the change of period of  pulsars   imposing that $\Lambda\gtrsim 1$ MeV. For such a large value of the coupling scale and $\beta={\cal O}(1)$ we can have $\epsilon_V \gtrsim 1$ when
$
p\lesssim (\frac{G_N M}{\Lambda^2})^{1/3} \sim 10^{-8}{\rm m}
$
which is obviously non-sensical. On the other hand, no bound on $\Lambda$ are known in the environment of White Dwarfs. As for all effective description using effective interaction suppressed by dimensionful scales, the value of $\Lambda$ could depend on the environment, i.e. on the energy scales in a particular setting. Large kicks for White Dwarf binaries could be envisaged and scalar effects tested provided $\Lambda$ is relaxed in WD environments and does not have to comply with the NS bound. A more fundamental model where the coupling of the scalar field to WD and NS would be obtained from first principle would be required to see if this possibility could be envisaged. This is left for future work.

\section{Conclusions}
\label{sec:conc}
Light scalar fields coupled to matter may be at the origin of 
phenomena like dark energy and dark matter. In this paper, we have considered the gravitational effects induced by conformal and disformal couplings of light scalars on binary system. More precisely, we have focused on hyperbolic orbits and the potential memory effects  and kicks that scalars may induce. First of all and contrary to GR, the link between the linear memory, i.e. the remaining deformation, after the passage of a binary system in a hyperbolic orbit and the power spectrum emitted at vanishing frequency is not direct. In particular, we find that only the conformal interaction leads to a contribution to the power spectrum for vanishing frequency. Moreover the scalar monopole  does not contribute and only the quadrupole is relevant. This is complemented by a similar result for the linear memory, i.e. the disformal interaction plays no role and only the conformal one leads to a non-vanishing effect. This asymmetry between the two types of scalar interactions is striking and further investigation is certainly required to unravel its origin. In particular, the role of asymptotic symmetries should be understood, see for instance \cite{AA} in the GR context. {The absence of disformal memory effect deserves further study. In particular, the role of asymptotic symmetries in the dual formulation of \cite{Seraj:2021qja} could lead to a better understanding of this result. The fact that scalars do not induce non-linear memory could also be a clue \cite{Tahura:2020vsa,Hou:2020xme,Heisenberg:2023prj}. We leave this to further investigation.}

Although the disformal interactions do not lead to linear memory, they give rise to kicks of  binary systems, i.e. the velocity of the centre of mass of a binary system is affected by the emission of scalar waves close to the point of closest encounter. This may result in a significant effect provided the coupling $\Lambda$ characterising the strength of the disformal interaction is lower than the bound provided by the dynamics of binary pulsars. If this bound can be relaxed the WD's in hyperbolic encounters may lead to observable effects. 
In particular, the concomitant observation of both memory effects and kicks for hyperbolic orbits would allow one to deduce the values of the conformal and disformal couplings. If light scalar fields become the natural candidates for dark energy or dark matter, their couplings to matter will become crucial for the confirmation of their existence not only observationally but experimentally. Gravitational  tests such as the ones presented here may then become of practical relevance.

\section*{Acknowledgements}

We would like to thank L.Bernard, G.Faye, F.Nitti and L.Santoni for useful comments or discussions.

\appendix

\section{Mean anomaly for hyperbolic orbits}
\label{app:Newton}

For hyperbolic Newtonian orbits $e>1$, integration of Eq.~(\ref{eq:dotp}) gives
\begin{equation}
t(\eta) = \nb (e\sinh \eta - \eta)
    \label{eq:tetaap}
\end{equation}
where 
\begin{eqnarray}
 \nb &=& 
   \sqrt{\frac{a^3}{G_Nm} }
    \label{eq:nbdef}
    \\
      a &=& \frac{p}{e^2-1}.
 \label{eq:adef}
\end{eqnarray}
and the `mean anomaly' $\eta$ satisfies
\begin{equation}
 \tanh (\eta/2) =  \sqrt{\frac{e-1}{e+1}}   \tan(\phi/2).
\label{eq:etadefap}
\end{equation}
 Eq.~(\ref{eq:r}) then becomes
\begin{equation}
    r(\eta) =a(e\cosh \eta - 1),
\end{equation}
while straightforward manipulations of (\ref{eq:etadefap}) give
\begin{eqnarray}
    \cos \phi &=& \frac{e - \cosh \eta}{e\cosh \eta - 1},
    \label{eq:cosphi}
    \\
    \sin \phi &=& \sqrt{e^2-1} \frac{\sinh \eta}{e\cosh \eta - 1}.
    \label{eq:sinphi}
\end{eqnarray}
Thus from Eq.~(\ref{eq:vecr}) $\vec{n} = \frac{a}{r(\eta) }\vec{m}$ with
\begin{equation}
    \vec{m} = \big( e - \cosh\eta, \sqrt{e^2-1} \sinh\eta \big).
\end{equation}

\section{Energy spectrum $P(\omega)$ and Hankel functions}
\label{app:spectrum}

An integral representation of  Hankel functions of the first kind is (see e.g.~\cite{GR}, section 8.421) 
\begin{eqnarray}
    H_\nu^{(1)}(xz) &=& -\frac{i}{\pi}e^{-i\nu \pi/2}  z^\nu \int_0^\infty dt \; t^{-\nu - 1} \exp\left[ \frac{ix}{2} \left( t + \frac{z^2}{t}\right) \right]
    \label{eq:H1def2}
\end{eqnarray}
with ${\rm arg}(z)=\pi/2, x>0$ and $-1 < \Re(\nu) < 1$.
Take $z^2 = -1$ so $z=i=e^{i\pi/2}$ (satisfying arg$(z)=\pi/2$).
Changing variables to $t=e^{-\eta}$, with $-\infty < \eta <\infty$ leads to
\begin{equation}
    H_\nu^{(1)}(ix) =  -\frac{i}{\pi} \int_{-\infty}^{\infty} d\eta e^{\nu \eta - ix \sinh \eta}.
\end{equation}
Now let $q=ix$ which recovers
Eq.~\eqref{eq:H1def}. We also use the notation introduced in Sec.~\ref{sec:scalar-memory}, namely $H_\nu^{(1)}(ix) =H_\nu^{(1)}(q) \equiv Z_\nu(q)$. As shown after Eq.~\eqref{here}, $\tilde{I}_{ij}$ is expressed in terms of $Z_\nu(q)$ with $q=ie\tw$.

After tedious calculations starting from Eq.~\eqref{here}, the diagonal components of the quadrupole tensor are given by
\begin{eqnarray}
    \tilde{I}^{kk}(\omega) &=& \left(\frac{\mu a^2}{3\omega_c} \right) i \pi \big[  \alpha_{3}^{kk} (Z_{i\tw + 3}(q) +Z_{i\tw - 3}(q)) 
 +\; \alpha_{2}^{kk} \left(Z_{i\tw + 2}(q) +Z_{i\tw - 2}(q)\right) 
  \nonumber \\ && \qquad\qquad\quad
 +\; \alpha_{1}^{kk} (Z_{i\tw + 1}(q) +Z_{i\tw - 1}(q)) 
 +\;\alpha_{0}^{kk} Z_{i\tw } (q) 
    \big],
    \label{eq:Iqq}
\end{eqnarray}
whereas the only non-zero off-diagonal component is
\begin{equation}
    \tilde{I}^{12}(\omega) 
    =\left(\frac{\mu a^2}{\omega_c} \right) i \pi\sqrt{e^2-1}  \left[  \alpha_{3}^{12} (Z_{i\tw + 3}(q) - Z_{i\tw - 3}(q)) 
+ \alpha_{2}^{12} (Z_{i\tw + 2}(q) -Z_{i\tw - 2}(q)) 
+ \alpha_{1}^{12} (Z_{i\tw + 1}(q) -Z_{i\tw - 1}(q)) 
    \right].
    \label{eq:tI12}
\end{equation}
For the diagonal terms, the different coefficients are given by
\begin{eqnarray}
    \alpha_{3}^{11} = -\frac{e}{8}(e^2-3)\, ,\qquad & \alpha_2^{11} = -\frac{3}{4}(e^2+1)\, ,& \qquad \alpha(r)^{11} = \frac{3e}{8}(3e^2+7) \, ,\qquad \alpha_0^{11} = \frac{1}{2}(9e^2+1)\, , \\
     \alpha_{3}^{22} = \frac{e}{8}(2e^2-3)\, , \qquad & \alpha_2^{22} = \frac{3}{4} \, , &  \qquad  \alpha(r)^{22} = -\frac{3e}{8}(2e^2+3) \, ,\qquad   \alpha_0^{22} = 3e^2-\frac{1}{2},
\end{eqnarray}
and the tracelessness of $I^{pq}$ implies 
\be
\alpha_{p}^{33} = -(\alpha_p^{11} +\alpha_p^{22}).
\ee
For the non-diagonal term 
\begin{equation}
    \alpha_3^{12} = -\frac{e}{8}, \qquad \alpha_2^{12} = \frac{e^2+1}{4},\qquad \alpha(r)^{12} = -\frac{5e}{8}. 
\end{equation}
The spectrum for hyperbolic orbits is then determined by substituting Eqs.~(\ref{eq:Iqq}) and (\ref{eq:tI12}) into Eq.~(\ref{eq:Pomega}), namely
\begin{equation}
    \pgw(\omega) = \frac{G_N}{5 \pi c^5} \left[ \omega^3 \tilde{I}^{ij}(\omega)\right]\left[ \omega^3 \tilde{I}_{ij}^*(\omega) \right].
\label{eq:Pomegabis}
\end{equation}

\section{The spectrum for vanishing frequency, $P(0)$}
\label{app:P0}

A check the results of Appendix \ref{app:spectrum} is to determine $\pgw(0)$, and then verify it agrees with the result obtained from the memory effect given in Eq.~\eqref{eq:P0}. 

Eqs.~(\ref{eq:Iqq}) and (\ref{eq:tI12}) can be simplified using the recursion relations for Hankel functions \cite{GR},
\begin{eqnarray}
    Z_{\nu+1}(q)+Z_{\nu-1}(q) &=& \frac{2\nu}{q} Z_\nu(q) \\
    Z_{\nu+1}(q)-Z_{\nu-1}(q) &=& 2 Z'_\nu(q)
\end{eqnarray}
where $Z'(q)=dZ/dq$. Recalling that $\nu=i\tw$, $q=ie\tw$, the relevant terms in $\tilde{I}^{kk}$ and $\tilde{I}^{12}$ become
\begin{eqnarray}
    Z_{i\tw+1}(q)+Z_{i\tw-1}(q) &=& \frac{2}{e} Z_{i\tw}(q)
    \label{eq:r1diag}
    \\
     Z_{i\tw+1}(q)-Z_{i\tw-1}(q) &=& 2Z'_{i\tw}(q)
     \label{eq:r1nondiag}
\end{eqnarray}
which in the zero frequency limit scale as \cite{GR}
\begin{eqnarray}
    Z_{i\tw}(q=ie\tw) &\underset{\tw \rightarrow 0}{\longrightarrow} &\frac{2i}{\pi} \ln(\tw e)
    \label{eq:limitZ}
    \\
     Z'_{i\tw}(q=ie\tw) &\underset{\tw \rightarrow 0}{\longrightarrow} &\frac{2}{\pi e \tw}.
     \label{eq:limitZ'}
\end{eqnarray}
Thus the combination of Hankel functions appearing in $\tilde{I}^{12}$ (namely Eq.~\eqref{eq:r1nondiag}) dominates over that appearing in the diagonal components (namely Eq.~\eqref{eq:r1diag})  in the zero frequency limit.

Similarly for $m=2,3$ the relevant recursion relations are
\begin{eqnarray}
     Z_{\nu+m}(q)+Z_{\nu-m}(q) &=& a_m Z_\nu(q) + b_m  Z'_\nu(q),
     \\
      Z_{\nu+m}(q)-Z_{\nu-m}(q) &=& c_m Z_\nu(q) + d_m  Z'_\nu(q)
\end{eqnarray}
where for $m=2$
\begin{eqnarray}
    a_2 = 2 \left(\frac{2\nu^2}{q^2}-1\right) = 2 \left(\frac{2}{e^2}-1\right), && 
    b_2 = - \frac{4}{q} = -\frac{4}{i e \tw},
    \\
    c_2 = -4 \frac{\nu}{q^2} =\frac{4i}{e^2 \tw},\qquad \qquad \qquad \; \;
   &&
    d_2 = \frac{2 \nu}{q} =\frac{2}{e},
\end{eqnarray}
and for $m=3$
\begin{eqnarray}
    a_3 & = & \frac{8\nu^3}{q^3} + 2\left(\frac{8}{q^3} - \frac{3}{q}\right)\nu = \frac{2}{e}\left(\frac{4}{e^{2}}-3\right)-\frac{16}{\tw^{2} e^{3}},\\
    b_3 &=& -\frac{24\nu}{q^2}= \frac{24 i}{e^{2}\tw },
    \\
    c_3 &=& -2 \left(-\frac{11}{q^{3}}-\frac{1}{q^{2}}\right) \nu^{2}-\frac{6}{q}-2 = 2\left(\frac{1}{e^{2}}-1 \right) + \frac{1}{\tw} \frac{2i}{e^3}\left(3e^2-11 \right),  \\
    d_3 &=& -8\frac{\nu^2}{q^2} - \frac{16}{q^2} + 2 = 2\left(1-\frac{4}{e^{2}} \right) + \frac{16}{\tw^2 e^2}.
    \label{eq:d3}
\end{eqnarray}

The power spectrum at zero frequency is determined by (see Eq.~(\ref{eq:Pomegabis}))
\begin{equation}
    \underset{\omega \rightarrow 0}{\lim } \omega^3 \tilde{I}_{ij}(\omega).
\end{equation}
Obviously the only term which will give a non-zero contribution is $d_3 Z'_\nu$ and in particular from Eqs.~\eqref{eq:limitZ'} and \eqref{eq:d3}
\begin{equation}
\lim_{\tw \rightarrow 0} d_3 Z'_\nu  =    \frac{32}{\pi \tw^3 e^3}.
\end{equation}
This is a part of $\tilde{I}_{12}$, so going back to (\ref{eq:Iqq}) and (\ref{eq:tI12}),
\begin{eqnarray}
\lim_{\omega \rightarrow 0}   \omega^3 |\tilde{I}^{kk}|&=&0,
\\
     \lim_{\omega \rightarrow 0}   \omega^3 |\tilde{I}^{12}| &=&
     4 {\mu}a^2 \omega_c^2 \frac{\sqrt{e^2-1}}{e^2}
\end{eqnarray}
which agrees with \eqref{eq:I120}.
Substituting into Eq.~\eqref{eq:Pomegabis}, together with the definitions $a=p/(e^2-1)$ and $\omega_c^2=G_Nm/a^3$, thus gives
\begin{eqnarray}
    \pgw(\omega=0) &=& 
      2 \frac{G_N}{5 c^5 \pi}16\mu^2 \left(\frac{G_N m}{p}\right)^2 \frac{{(e^2-1)^3}}{e^4} 
      \label{eq:agg}
\end{eqnarray}
which agrees entirely with Eq.~(\ref{eq:P0}) calculated through the linear memory effect.
This also coincides with the result in \cite{Caldarola:2023ipo} (though our power spectrum is divided by $2\pi$ compared theirs).
Thus as in the main text we can write
\begin{equation}
     \pgw(\omega)= \pgw(0) {f}(\tw,e)
\end{equation}
where $\pgw(0)$ is given in \eqref{eq:agg}. The dimensionless ${f}_{\rm GW}(e,\tw)$ equals 1 at zero frequency, and it is plotted in figure \ref{fig:dimensionless-power-GR}. See section \ref{subsec:memory} for further discussion.

\section{Calculation of the Scalar momentum force}
\label{app:derivr}

The following identities will be useful. From Newton's law to lowest order $\vec{a} = -\Geff  m r^{-3} \vec{r}$, one has
\begin{eqnarray}
\frac{d^2}{dt^2}\left(\frac{1}{r}\right) &=& \frac{3}{r^5} (\vec{r}\cdot \vec{v})^2 - \frac{v^2}{r^3} + \frac{\Geff  m}{r^4}
\label{eq:rinverse2}
\\
\frac{d^3}{dt^3}\left(\frac{1}{r}\right) &=& -\frac{15}{r^7} (\vec{r}\cdot \vec{v})^3 + \frac{9}{r^5}(\vec{r}\cdot \vec{v}) v^2 - \frac{8 \Geff  m}{r^6}  (\vec{r}\cdot \vec{v})
\label{eq:rinverse3}
\\
    \frac{d^4}{dt^4}\left(\frac{1}{r}\right) &=& \frac{75 \Geff  m}{r^8} (\vec{r}\cdot \vec{v})^2 - \frac{90}{r^7} v^2 (\vec{r}\cdot \vec{v})^2 + \frac{15\cdot 7}{r^9}(\vec{r}\cdot \vec{v})^4 
    \nonumber
    \\
    &&+ \frac{9}{r^5}(v^2)^2 - \frac{17 \Geff  m}{r^6} v^2 + \frac{8}{r^7} (\Geff  m)^2
    \label{eq:rinverse4}
\end{eqnarray}

The scalar momentum force is given in Eq.~\eqref{eq:momentumradiated} namely 
\be 
\frac{dP^i}{dt}=- \frac{4 G_N}{3}    \dot I_{\phi}\ddot I^{i}_\phi- {2}\frac{4G_N}{15}\ddot I_{j\phi}\dddot I^{ij}_\phi {-} \cF^i.
\ee
Working to order $\Lambda^{-2}$, only the second term in (\ref{eq:Iphi}) will contribute to $\dot{I}_\phi$ with
\begin{equation}
\dot{I}_\phi = - \frac{8\beta G_Nm\mu}{3r^3} (\vec{r}\cdot\vec{v}).
\end{equation}
From (\ref{eq:Iphii}),
\begin{eqnarray}
\dot I_{j\phi}&=&- \tilde{\xi} \left(r_j \dddot{(1/r)} + v_j \ddot{(1/r)} \right)
\\
 \ddot I_{\phi}^i&=&- \tilde{\xi} \left(d_1 r^i + d_2 v^i\right) = - \tilde{\xi} \left(e_1 n^i + e_2 \lambda^i\right)
 \label{eq:IddotIi}
\end{eqnarray}
where
\begin{eqnarray}
\tilde{\xi} \equiv \frac{4\beta \Geff \mu m}{\Lambda^2} \Delta &&
\label{eq:txidef}
\\
      d_1 \equiv \ddddot{(1/r)} - \frac{\Geff mb}{r^3} \ddot{(1/r)}
       & \qquad & d_2 \equiv 2 \dddot{(1/r)}
       \nonumber\\
      e_1 \equiv d_1 r + d_2 \dot{r}
        & \qquad &
        e_2 \equiv d_2 r \dot \phi.
         \label{eq:e12def}
\end{eqnarray}
Using the different derivatives of $1/r$ given in Eqs.~\eqref{eq:rinverse2}-\eqref{eq:rinverse4}, 
the contribution to the scalar force from the monopole-dipole contribution is found to be
\begin{equation}
\left( {\cal{F}}_\phi^{m-d} \right)^i \equiv \frac{4G_N}{3}  \dot I_{\phi} \ddot I^{i}_\phi =\frac{4G_N}{3}   \left( \frac{8 }{3}\right)\tilde{\xi}(\beta G_N \mu m) \left( \frac{\dot{r}}{r^2}\right) \left[e_1 n^i + e_2 \lambda^i \right].
\label{eq:annecy1}
\end{equation}

Regarding the dipole-quadrupole contribution to the scalar force, using results in \cite{poisson_will_2014} (see equation (12.77), divided by $\eta m$)
\begin{eqnarray}
    \dddot{I}_\phi^{ij} = \beta \mu \left( \frac{2 \Geff m}{r^2}\right) \left[ \dot{r} n^i n^j + 2r\dot{\phi}(n^i \lambda^j + n^j \lambda^i) \right] .
\end{eqnarray}
Thus from Eq.~\eqref{eq:IddotIi} it follows that 
\begin{equation}
 \left({\cal{F}}_\phi^{d-q}\right)^i\equiv \frac{8G_N}{15} \ddot I_{j\phi}\dddot I^{ij}_\phi  = -\frac{8G_N}{15} 2 \tilde{\xi} (\beta \Geff \mu m ) \left( \frac{1 }{r^2}\right) \left[(e_1\dot{r} + 2 r \dot{\phi}e_2) n^i + 2 r \dot{\phi} e_1 \lambda^i \right].
 \label{eq:annecy2}
\end{equation}
Since we are working to order $\beta$, we set $\Geff=G_N$ in the previous equation as well as in $\tilde{\xi}$. Hence one finally has
\begin{equation}
\vec{{\cal{F}}}_\phi =  \vec{{\cal{F}}}_\phi^{m-d} +\vec{{\cal{F}}}_\phi^{d-q} = +\frac{Y}{r^2} \left[ (21\dot{r} e_1 -18 r\dot{\phi} e_2)\vec{n} + (30\dot{r}e_2 - 18 r \dot{\phi} e_1) \vec{\lambda} \right]
\label{eq:annecy3}
\end{equation}
where
\begin{equation}
    Y = \frac{16 G_N \tilde{\xi} (\beta G_N \mu m)}{135}
\end{equation}
Thus to leading order, using $\vec{n}$ and $\vec{\lambda}$ in Eqs.~\eqref{eq:vecr}-\eqref{eq:vecv}, the $x$ and $y$ components are
\begin{eqnarray}
 \left[\vec{{\cal{F}}}_\phi\right]_x &=&
\frac{3e Y}{p} \left(\frac{G_Nm}{p^3}\right)^{5/2}
(1+\cos\phi)^6 \sin\phi
 \left[ 
1095 e^3 \cos^5\phi  + 1614 e^2\cos^4\phi + e \cos^3\phi (-1362 e^2 + 667)\right. 
\nonumber
\\
&& \left.  + (-1828 e^2 + 64)\cos^2\phi + e\cos\phi(276 e^2 - 738 ) + 192 e^2 - 104
\right]
\label{eq:Fphix}
\\
\left[ \vec{{\cal{F}}}_\phi \right]_y &=& \frac{3e Y}{p} \left(\frac{G_Nm}{p^3}\right)^{5/2}
(1+\cos\phi)^6
 \left[ 
-1095 e^4\cos^6\phi  - 1614 e^3\cos^5\phi  + (1707 e^4 - 667e^2)\cos^4\phi 
\right. 
\nonumber
\\
&& \left. + (1964 e^3 - 64 e)\cos^3\phi + (-750 e^4 + 495 e^2)\cos^2\phi + (-524 e^3 - 26e)\cos\phi + 84 e^4 - 26 e^2 - 12
\right]
\end{eqnarray}

Following the same procedure as in section \ref{subsec:kicks}, the shift of the CM velocity due to the scalar field again vanishes in the $x$-direction. Its component in the $y$ direction can be written in terms of $v_{cm}$ defined in \eqref{eq:vcmGR} leading to 
\begin{equation}
      \Delta V^y_\phi =
   v_{cm} 
   \left(\frac{\beta^2 G_Nm}{\Lambda^2 p^3} \right) f_\phi^y(e)
\end{equation}
where
\begin{eqnarray}
    f^y_\phi &=& 
    \frac{1}{41760e } \left[
   \sqrt{e^2-1} (1783057 e^8 + 10915688 e^6 + 9108564 e^4 + 942560 e^2-2144) \right.
    \nonumber
    \\
    && \qquad \left. + 150255 e^2\arccos(-1/e)\left(
    e^8 + \frac{14090}{477}e^6+ 
    \frac{38480}{477}e^4+
    \frac{2016}{53}e^2
+\frac{1024}{477}    \right)
    \right]
    \label{eq:fphiy}
\end{eqnarray}
Thus this {\it increases} the kick velocity relative to the GR case. Also notice that the terms scale with larger powers of $e_\Lambda$ (here $e^8$ rather then $e^4$ in GR) and hence, despite the small coefficient $\epsilon$ in front, these terms could become significant for large eccentricities.

\bibliography{main}

\end{document}